\documentclass[letterpaper,twocolumn,10pt]{article}
\usepackage{usenix-2020-09}

\usepackage{tikz}
\usepackage{amsmath}
\usepackage{enumitem}
\usepackage{tabularx}
\usepackage{multirow}
\usepackage{amssymb}
\usepackage{hyperref}
\newcolumntype{Y}{>{\centering\arraybackslash}X}
\usepackage{xurl}
\usepackage{siunitx}
\usepackage{lipsum}

\newcommand\blfootnote[1]{
  \begingroup
  \renewcommand\thefootnote{}\footnote{#1}
  \addtocounter{footnote}{-1}
  \endgroup
}

\newcommand{\newtext}[1]{#1}

\begin{document}

\date{}

\title{\Large \bf Understanding and Improving Usability of Data Dashboards for \\ Simplified Privacy Control of Voice Assistant Data\\(Extended Version)}

\author{
{\rm Vandit Sharma}\\
Indian Institute of Technology Kharagpur\\
vanditsharma@iitkgp.ac.in
\and
{\rm Mainack Mondal}\\
Indian Institute of Technology Kharagpur\\
mainack@cse.iitkgp.ac.in
}

\maketitle

\begin{abstract}

\noindent Today, intelligent voice assistant (VA) software like Amazon's Alexa, Google's Voice Assistant (GVA) and Apple's Siri have millions of users. These VAs often collect and analyze huge user data for improving their functionality. However, this collected data may contain sensitive information (e.g., personal voice recordings) that users might not feel comfortable sharing with others and might cause significant privacy concerns. To counter such concerns, service providers like Google present their users with a personal data dashboard (called `My Activity Dashboard'), allowing them to manage all voice assistant collected data.
\newtext{However, a real-world GVA-data driven understanding of user perceptions and preferences regarding this data (and data dashboards) remained relatively unexplored in prior research.}

\newtext{To that end, in this work we focused on Google Voice Assistant (GVA) users and investigated the perceptions and preferences of GVA users regarding data and dashboard while grounding them in real GVA-collected user data.}
Specifically, we conducted an 80-participant survey-based user study to collect both generic perceptions regarding GVA usage as well as desired privacy preferences for a stratified sample of their GVA data. We show that most participants had superficial knowledge about the type of data collected by GVA. Worryingly, we found that participants felt uncomfortable sharing \newtext{a non-trivial 17.7\%} of GVA-collected data elements with Google. The current My Activity dashboard, although useful, did not help long-time GVA users effectively manage their data privacy. \newtext{Our real-data-driven study found that showing users even one sensitive data element can significantly improve the usability of data dashboards.} To that end, we built a classifier that can detect sensitive data for data dashboard recommendations with a 95\% F1-score and shows 76\% improvement over baseline models.
\blfootnote{\noindent This extended version of our USENIX Security ’22 paper includes appendices for interested readers.}

\end{abstract}

\section{Introduction}

Voice assistants like Google's voice assistant (GVA), Amazon's Alexa, Microsoft's Cortana, or Apple's Siri are extremely popular today as they are well equipped to perform multiple tasks on users' voice requests (e.g., searching the internet, calling a friend, or playing music). However, these voice assistants also collect and analyze a lot of user data (e.g., timestamps, audio recordings, transcripts, etc.) to improve their infrastructure across multiple devices (e.g., in both smart speaker and smartphone). Unfortunately, this data can lead to a huge possible privacy nightmare since the voice assistant might be used in private situations.
E.g., GVA collects three types of potentially sensitive data---audio clips of conversations, transcripts of conversations, and the ambient location of use. We refer to individual records of these three data types as data elements in this paper.

\newtext{In this study, we take Google voice assistant (GVA) as our experimental testbed}.  Previous studies on understanding user perceptions and preferences for data collection by voice assistants (such as~\cite{malkin2019attitudes, tabassum2019always}) have mainly focused their attention on smart speaker users. However, recent reports~\cite{va2018adoption, voicebot2020smartphones} have highlighted the significantly greater popularity of smartphone-based voice assistants over their smart speaker counterparts. Intuitively, smartphones are easier to use in more contexts than smart speakers, multiplying potential privacy problems. \newtext{To that end, the exact same GVA software runs in both Google smart speakers and Android smartphones, effectively aggregating data from both. So, we focus on GVA users and conduct a real user-data driven study to uncover user perceptions regarding GVA-collected data.}

Specifically, to counter this problem of sensitive data collection, service providers like Google often provide a dashboard to the users showcasing their GVA collected data (Google's My Activity dashboard). \newtext{We noted that the dashboard includes data from both smart speakers and smartphones without differentiating markers}. However, the efficacy of these data dashboards for controlling privacy in the GVA context is not well-understood.
To that end, we unpack user perceptions and preferences regarding data collection by GVA as well as data dashboards through a two-part survey-based user study.
\newtext{Our overall goal is to assess the usefulness/efficacy of data dashboards. We specially focus on the context of data collected by voice assistants (in smart devices) and investigate the efficacy of these dashboards to enable the privacy goals of users in that context.}

\newtext{Our research questions (RQs), as stated below, are designed to unravel (i) whether data dashboards can indeed facilitate a better understanding of what (possibly sensitive) data VAs collect, and (ii) the particular helpful (or not so helpful) aspects of data dashboards from a user-centric view. Our RQs also investigate how to improve the usability of these data dashboards. In this study, we particularly contextualize our RQs with our focus on GVA. We selected GVA primarily because of the huge user base (boosted by the inclusion of GVA in all Android smartphones). Even though our choice of GVA poses some limitations, (e.g., GVA userbase and dashboard might not necessarily represent all VA users or dashboards), our approach is still useful---findings from our study answer broader questions about helpfulness of data dashboards in general and take a step forward towards improving their usability.}

\vspace{1mm}
\noindent \textbf{RQ1}- How frequently do Android users leverage GVA? What is their GVA usage context?

Most (72.5\%) of our participants had been using GVA for around two years or more. 73.75\% of participants used GVA frequently, at least a couple of times a week in home, office and car. For the median participant, GVA collected 837.5 data elements.
The context of using GVA ranged from getting information to entertainment.

\vspace{1mm}
\noindent \textbf{RQ2}- What are the user perceptions regarding the data collection and storage practices of GVA? What is their ideal access control preference for Google relative to their social relations for accessing GVA data?

Although the majority (78.75\%) of participants were aware that Google collects and stores some form of data, around 40\% of users were unclear about the type of data (e.g., audio clips) being collected, signifying superficial knowledge.
Interestingly, statistical analysis shows that relative to various social relations (proxemic zones~\cite{hall1966hidden}), participants considered Google mainly as a public entity.

\vspace{1mm}
\noindent \textbf{RQ3}- Do users desire to restrict access of Google to GVA collected specific data elements? Do these access preferences correlate with the data element class or the medium of the data collection?

Participants wanted to restrict Google's access for 121 (18.08\%) out of 669 audio clips and 61 (17.03\%) out of 358 transcripts presented in our survey, a non-trivial fraction of all collected data. They had similar preferences for data collected by smartphones and smart speakers but felt significantly less comfortable viewing data elements where they did not know or could not recall the device through which it was collected. There were no significant differences in user privacy preferences for data elements from different control and possibly sensitive classes (prepared from previous work~\cite{malkin2019attitudes, bentley2018usage, yuting2019ipa}), suggesting the inherent complexity of finding sensitive data.

\vspace{1mm}
\noindent \textbf{RQ4}- Do data dashboards help users to control the privacy of their GVA data? Can we further assist users by automated means to improve their privacy-preserving behavior by improving the data dashboard? How?

50\% of our participants did not know about the Google-provided My Activity data dashboard. Most participants found the dashboard easy to use; however, more long-time GVA users expressed a need for assistance in using the dashboard, suggesting a lack of effectiveness for larger amounts of data. Showing users even one sensitive data element collected by GVA using our simple class-based sensitive content detection system made them highly (80\%) likely to control their collected data. This suggests that assisting dashboard users through automated means might improve their privacy-preserving behavior. We took the first step in this direction by exploring an Machine learning (ML)-assisted human-in-the-loop (HITL) based design for data dashboards. We show that it is possible to create Machine learning-based systems to recommend sensitive content with more than 95\% F1-score showing a concrete, feasible direction to improve data dashboards.
\newtext{We note that, although we used GVA as our experimental testbed, our findings
regarding the efficacy of data dashboards as well on improving data dashboards could be extended to contexts concerning other VAs. For example, our results show that dashboards are indeed useful for tracking VA-collected data. However, there also is a need for automated assistance in using the dashboards, notably for long-term users to control the privacy of large amounts of accumulated data. These findings are potentially useful for designing improved data dashboards for any VAs.}

The rest of the paper is organized as follows- The background and related work in Section~\ref{sec:background}. Our methodology is explained in Section~\ref{sec:methodology}. We describe the data analysis in Section~\ref{sec:analysis}. The survey results are presented in Section~\ref{sec:results}. In Section~\ref{sec:recommendation}, we explore ML as a possible improvement to recommend sensitive data elements in data dashboards. Finally, we conclude the paper in Section~\ref{sec:conclusion}.

\section{Background and Related Work}
\label{sec:background}

\begin{figure}[!t]
\centering
\includegraphics[width=\linewidth]{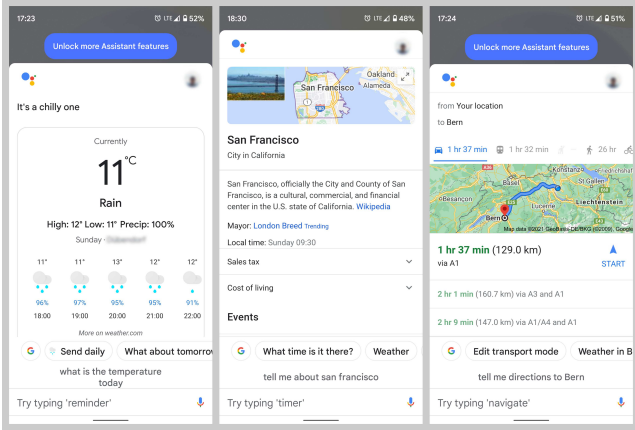}
\caption{GVA interface on an Android smartphone.}
\label{fig:assistant}
\end{figure}

\noindent \textbf{GVA capabilities and usage:} GVA is an intelligent voice-activated assistant software that Google introduced in May 2016~\cite{gva2016launch}. It allows users to perform a variety of actions such as getting local information, playing media, performing a Google search, managing tasks, and more~\cite{gva2021all} through simple voice-based commands. GVA supports cross-device functionality and is available on a wide range of devices such as smartphones, tablets, smartwatches, TVs, headphones, and more~\cite{gva2021all}. As of 2020, GVA is available on more than 1 billion devices, spread across 90 countries and supports over 30 languages. It has more than 500 million monthly active users~\cite{gva2020billion}, reflecting its immense popularity. \newtext{Figure~\ref{fig:assistant} shows a visual of the GVA interface and some functionalities.}

While most users of VAs use them on smartphones, tablets, and smart speakers~\cite{pew2017smartphones}, recent privacy studies have been paying increased attention to concerns surrounding the use of smart speakers~\cite{cision2020smartphones}, primarily because they are always listening devices. However, several works point out the popularity of VA residing in smartphones which can capture more diverse contexts and potentially private data. This work focuses on GVA, which runs on both smart speakers and smartphones.

\vspace{1mm}
\noindent \textbf{Privacy concerns with VAs:} Privacy concerns surrounding voice assistants have been studied extensively. Several researchers have proposed different approaches to launch privacy attacks against voice assistants~\cite{zhang2017dolphin, alepis2017monkey, carlini2016hidden, zhang2019attack}. Schonherr et al. explored the accidental triggering of voice assistants~\cite{schonherr2020unacceptable}, and Edu et al. conducted a detailed literature review of Smart Home Personal Assistants (SPA) from a security and privacy perspective~\cite{edu2020smart}. They highlighted several key issues such as weak authentication, weak authorization, profiling, etc. Edu et al. also studied various attacks on SPAs, suggested countermeasures, and discussed open challenges in this area. Courtney further summarized various privacy concerns associated with voice assistants~\cite{courtney2017careless}.

In recent years, there have been multiple instances of data leaks associated with voice assistants managed by prominent technology companies~\cite{gva2019leak, alexa2020history}. Such data leaks can be a huge cause of privacy concern since VA collected data can include sensitive data such as audio recordings, location data, etc. \newtext{Specifically, one interaction with GVA can lead to multiple data elements---audio clip, transcript, and location---depending on the controls set in Google-wide settings (e.g., Web \& App Activity control, which is turned on by default and enables Google to store transcripts, location, and other metadata for all interactions).} For instance, Kr{\"o}ger et al. discussed the threat of unexpected inferences (such as obtaining the speaker’s identity, personality, age, emotions, etc.) from audio recordings stored by microphone-equipped devices through voice and speech analysis~\cite{kroger2020inference}. The two major classes of entities that can cause privacy violations with collected VA data are (i) the technology companies who own the voice assistants and store data on their servers (e.g., Google, Amazon, etc.), and (ii) external third parties with access to collected VA data, upon which technology companies might rely to review collected data. These two entities comprise our threat model.

\vspace{1mm}
\noindent \textbf{Managing privacy of VA-collected data:} A possible alternative to prevent privacy violations is limiting and controlling the data collected by voice assistants. Over the years, researchers have proposed several techniques for this purpose, involving both hardware and software. Champion et al. developed a device called the Smart\({}^{\mbox{2}}\) Speaker Blocker to address the privacy and security concerns associated with smart speaker use~\cite{champion2019blocker}. The device functioned by filtering and blocking sensitive information from reaching the smart speaker's microphone(s). However, such an intervention cannot be used in the case of smartphone-based voice assistants since smartphones are portable devices. Vaidya et al. proposed another technique to limit privacy concerns (such as voice spoofing) by removing certain features from a user's voice input locally on the device~\cite{vaidya2019voice}. Qian et al. additionally presented VoiceMask, a robust voice sanitization and anonymization application that acts as an intermediate between users and the cloud~\cite{qian2017voicemask}. Earlier work also developed a user-configurable, privacy-aware framework to defend against inference attacks with speech data.

While these techniques may be effective in checking privacy concerns, a significant downside is that they modify the collected data, rendering it unusable by developers. This defeats the purpose of collecting data in the first place since voice assistant developers need user data to train better ML models and improve their services. Tabassum et al. presented this as a privacy-utility trade-off, suggesting the development of privacy frameworks that allow users to control the amount of data collected by the voice assistant (in exchange for possibly limited services)~\cite{tabassum2019always}. The survey conducted by Malkin et al. on understanding the privacy attitudes of smart speaker users also highlighted a demand for effectively filtering accidental recordings and sensitive topics~\cite{malkin2019attitudes}.
We take a first step in this direction by exploring a human-in-the-loop design to identify and recommend sensitive data to GVA users.

\vspace{1mm}
\noindent \newtext{
\textbf{Privacy dashboards:} Following up on the recommendations made by the Abramatic et al. for better user privacy control~\cite{abramatic2015privacy}, Irion et al. advocated the use of privacy dashboards as a practical solution to enhance user control for data collected throughout the online and mobile ecosystem, including platforms such as GVA~\cite{irion2017roadmap}. They also highlighted the potential of AI techniques and methods to users manage and enforce their privacy settings. In this area, Raschke et al. presented the design and implementation of a GDPR-compliant and usable privacy dashboard~\cite{raschke2018dashboard}. In fact, to that end, Feth et al. proposed generic requirement and quality
models to assist companies with developing privacy dashboards for different domains~\cite{feth2020requirement}. Our research is motivated by this prior work on the importance of privacy dashboards---we aim to uncover the efficacy and possible improvement of today's privacy dashboards by specifically focusing on a deployed system in our real-world data-driven study.}

\vspace{2mm}
\noindent \textbf{Google My Activity data dashboard:} The Google My Activity dashboard is the primary data privacy control provided by Google for all its products and services. It is a hub where users can see and modify all of the key information that Google has been collecting about them over the years~\cite{cnet2016dashboard}. Figure~\ref{fig:myactivity} (Appendix)
shows the user interface of the Google My Activity dashboard. Since GVA's launch in 2016, not much work has been done on studying user perceptions and the utility of such data dashboards to manage data privacy.

A recent and closely related study~\cite{farke2021dashboards} investigated user perceptions and reactions to the Google My Activity dashboard. Through a survey, this study showed that viewing the My Activity dashboard significantly decreases concern about Google’s data collection practices. However, the authors were unsure if the dashboard actually provided valuable assistance in reviewing the collected data and enforcing user privacy. The first part of our study partially revisits this work. However, we answer several additional questions about the perceptions and preferences of GVA users. We also answer some questions raised by this study, such as the effectiveness of the Google My Activity dashboard to enforce user privacy. We then present a possible solution to improve data dashboards---through recommending sensitive data elements to users. We also demonstrate a highly accurate proof-of-concept human-in-the-loop-based machine learning model for the same.

\section{Methodology}
\label{sec:methodology}

\begin{figure}[!t]
\centering
\includegraphics[width=0.9\linewidth]{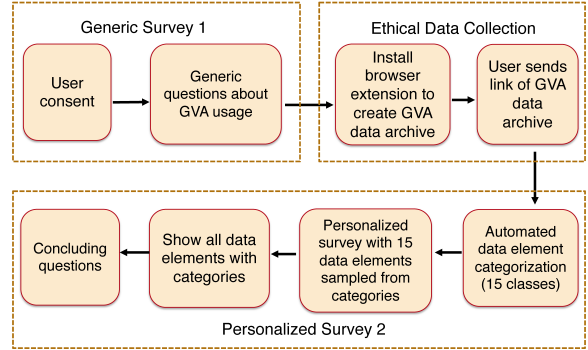}
\caption{Three key sections of our study---generic Survey~1, ethical data collection and personalized Survey~2.}
\label{fig:flow}
\end{figure}

\noindent We conducted a two-part survey-based user study to unpack perceptions and privacy preferences regarding GVA.

\subsection{Recruitment and Inclusion Criteria}

\noindent We deployed our study in the crowdsourcing platform Prolific Academic~\cite{prolific} during September 2020. We recruited 18+ years old US nationals with $>$95\% approval rating on Prolific. Additionally, we required that our participants primarily used an Android smartphone, used GVA more than once per month in the past year, and were willing to install our browser extension to share their GVA-collected data. We took the help of a short screening survey (Appendix~\ref{subsec:pre-screening})
which took less than a minute with Prolific-suggested compensation of \$0.11 to identify potential participants. Ultimately, we invited 249 participants (who satisfied our inclusion criteria) for participating in our actual two-part study. Survey 1 and Survey 2 of our study (seven days apart) took \newtext{a total of} 52 minutes and 42 minutes on average, respectively. We compensated participants who completed both the parts with \$12 (\$5 for part 1 and \$7 for part 2). In total, 80 participants completed both surveys (out of 119 who responded to our initial invitation). \newtext{This drop in the number of participants was potentially due to the task description and eligibility in the recruitment text.} \newtext{We consider the data from only these 80 participants in this paper as we wanted to combine data from both surveys (and, in effect, connect self-reported general perception from Survey 1 with the user feedback on real-world GVA-collected data in Survey 2).}

\subsection{Overview of Study Design}

\noindent Figure~\ref{fig:flow} summarizes our institutional ethics committee-approved study procedure. The study consisted of three main sections- (i) generic Survey 1, (ii) Ethical GVA data collection and (iii) personalized Survey 2.
First, participants were explained the study design as well as the exact data they needed to share.
The participants who gave us informed consent first took the generic Survey 1. This survey contained generic questions (instrument in Appendix~\ref{subsec:survey1})
regarding user knowledge and usage of Android smartphones as well as GVA. After completing Survey 1, participants installed a browser extension developed by us for ethical GVA data collection. Our extension worked entirely on the client-side and helped users create an archive of GVA data and upload it to their own Google account. Then, the participants manually shared a link to the online archive with us. Next, we leveraged an end-to-end fully automated pipeline to fetch participants' shared GVA data and processed the data in a secure computer. No researcher ever manually saw or analyzed the raw data. This processing phase identified possibly sensitive data elements collected by GVA. Then, within seven days of completing Survey~1 and sharing data, we invited the participants to return for a personalized Survey 2 (instrument in Appendix~\ref{subsec:survey2}).
In Survey~2, we elicited user perceptions of a stratified sample of these possibly sensitive data elements. \newtext{Since Survey~2 was generated programmatically for each participant using elements in their own GVA-collected data, we refer to it as a ``personalized survey''.} We also showed each participant all of their possibly sensitive data elements identified by our data processing pipeline in a personalized Google Drive folder (with named files and subfolders for categories) created by us. Finally, we asked the participants about the utility of automatically detecting sensitive GVA-collected data elements. Next, we explain each of the three sections of our study.

\subsection{Survey 1}

\noindent Our participants provided their online informed consent before starting Survey~1. In the consent form, we highlighted the purpose of our study, the specific data we would ask to share, and our privacy-preserving data collection and processing approach. Then, in Survey~1, we first asked participants some general questions to uncover their usage of Android smartphones and GVA. Next, drawing from earlier studies on
privacy concerns surrounding voice assistants, we designed a set of GVA usage scenarios to ground the user and uncover experiences with sensitive and even privacy-violating data collected by GVA~\cite{malkin2019attitudes, bentley2018usage, yuting2019ipa}. These scenarios ranged from ``Using inappropriate language'' and ``Using GVA in places with audible background sounds'' to ``Accidental activation of GVA'' (complete list in Table~\ref{tab:scenarios} of Appendix~\ref{subsec:basic}).
Then, our participants self-reported whether they recalled using GVA in these scenarios and their comfort in such contextual GVA usage. After this, participants responded to questions about their perceptions regarding GVA data collection (in general and under different transmission principles~\cite{apthorpe2018ci}), storage, and access, including a few questions specifically about Google My Activity dashboard. Finally, we concluded Survey~1 by asking questions related to general privacy attitudes.

\subsection{Ethical Collection of GVA \newtext{Data}}
\label{subsec:collection}

\noindent \newtext{Given the sensitive nature of the GVA-collected data elements, we wanted to collect it in the most ethical manner possible, as we will describe next. Our data collection protocol and analysis plan were thoroughly evaluated and approved by our Institutional Ethics Committee (equivalent to an IRB). Participants were briefed about the data collection process through the consent form at the beginning of the study.}

\vspace{1mm}
\noindent \textbf{Deciding on an ethical data collection protocol:} \newtext{We explored several options to collect GVA data ethically from users along with their downsides---a client-side data-analysis approach was infeasible due to the scale of data and computation, a Google password sharing approach encouraged oversharing private data, and approaching Google to analyze user data and performing our study could potentially be perceived as diminishing user agency. We finally asked our participants to use Google Takeout~\footnote{A Google service that enables users to export part or all of data elements stored in Google servers in an archive file~\cite{lifewire2020usingtakeout}.}, create an archive of \textit{only} GVA data by selecting specific options in the Google Takeout interface, and share the archive file with us after reviewing the data. We created a Firefox browser extension to facilitate data collection---(i) The extension automatically selected the right options in the Google takeout interface (in client browser) to create an archive with \textit{only} GVA data by choosing the right options in the Google Takeout interface. This approach diminished the chances of oversharing (e.g., chances of accidentally adding \newtext{all their emails}). (ii) The extension automatically selected the option provided in Google Takeout to create an archive in a user's own Google cloud storage~\footnote{\newtext{We took due consent to store and share this data in participants' personal cloud storage. The consent form is in Appendix~\ref{subsec:survey1}.
}} (associated with Google account). A participant shared their unique link with us to allow processing of their archive file.}

\vspace{1mm}
\noindent \textbf{Ensuring privacy of our collected data:} \newtext{In our protocol, participant GVA data could only be accessed using unique individual links
Moreover, we informed the participants that they could revoke access anytime,
All GVA-collected data was anonymous since it did not include any email or names of users. On receiving a link, an automated pipeline checked the validity of the data (using data type and folder structure of the shared data) and invited only participants with valid data for Survey~2. All data processing was automated (no manual exploration of raw GVA-data) and was done in password-protected computers accessible only to the researchers.}

\subsection{Survey 2}

\begin{table}[!t]
    \footnotesize
    \centering
    \begin{tabularx}{\linewidth}{|X|p{4.3cm}|p{1cm}|}
        \hline
       \textbf{Class} & \textbf{Description} & \textbf{med. \#}\\
        \hline
        audio-noise & Audio with high background noise & 3\\ \hline
        audio-non-bkgd & Audio with low background noise & 94\\ \hline
        audio-multi-spkr & Audio with multiple speakers & 36\\ \hline
        audio-non-gend & Audio with non-dominant gender speaker & 57\\ \hline
        audio-grammar & Audio with grammatical error & 9\\ \hline
        audio-non-eng & Audio with non-standard English word & 51\\ \hline
        audio-regret & Audio with regret word & 25\\ \hline
        audio-neg-sent & Audio with negative sentiment & 32\\ \hline
        audio-rand & Audio not in the above categories & -\\ \hline
        transcript-typo & Transcript with grammatical error & 5\\ \hline
        transcript-non-eng & Transcript with non-standard English word & 28\\ \hline
        transcript-regret & Transcript with regret word & 20\\ \hline
        transcript-neg-sent & Transcript with negative sentiment & 19\\ \hline
        transcript-rand & Transcript not in the above categories & -\\ \hline
        location & Location data (e.g., latitude-longitude) & 10\\
        \hline
    \end{tabularx}
    \caption{Description of fifteen classes from Survey~2 for classifying GVA collected audio clips, transcript and location data. The last column signifies the median number of data elements per user for our participants.}
    \label{tab:classes}
\end{table}

\noindent Our personalized Survey~2 primarily involved eliciting user reactions regarding specific data elements collected and stored by GVA. We start with our data processing pipeline to select data elements for Survey~2.

\vspace{1mm}

\noindent \textbf{Creating a classifier to categorize data elements}: We identified (and leveraged in Survey~1) a set of privacy-violating scenarios where according to earlier work, potentially sensitive data might be collected by GVA~\cite{malkin2019attitudes, bentley2018usage, yuting2019ipa}. We analyzed these scenarios to create twelve classes that encompass potentially sensitive GVA-collected data elements. These data elements were broadly of two types---GVA-collected audio clips and transcripts of the commands given to GVA. Aside from these twelve classes, we considered three additional classes---a separate class ``location'' for location data, and two separate classes (``audio-rand'' and ``transcript-rand'') which identify audio clips and transcripts not belonging to any of the twelve classes and act as a baseline for data elements. These total fifteen classes are presented in Table~\ref{tab:classes}.

We then created automated classifiers to categorize data elements in each of these classes for each user. These classifiers primarily relied on off-the-shelf signal processing (e.g., measuring Signal to Noise Ratio or detecting the number and gender of speakers) and NLP techniques (finding a grammatical error, non-English word or negative sentiment). We created one binary classifier for each of the above-mentioned twelve classes in Table~\ref{tab:classes} (aside from ``location'', ``audio-rand'' and ``transcript-rand'' classes). These classifiers categorized GVA-collected audio clips and transcripts into one or more of these classes. The motivation and detailed description of each classifier is in
Appendix~\ref{subsec:basic}.

\vspace{1mm}
\noindent \textbf{Selecting individual data elements for Survey~2:} Once we classified each data element into one or more categories (with the help of classifiers) from Table~\ref{tab:classes}, we used a stratified sampling approach. In short, we randomly selected one data element from each category (without replacement) and used them to create the Survey~2 questionnaire. We also created a personalized Google Drive folder for each participant to review in Survey~2. The folder contained all possibly sensitive data elements found in their GVA data, arranged in thirteen respectively named files and subfolders (excluding ``audio-rand'' and ``transcript-rand'' categories). Note that our pipeline handled all of the above tasks automatically. Once personalized Survey~2 was generated, one researcher manually invited the corresponding participant (within seven days of data upload) to participate in Survey~2 via messaging on Prolific.

\vspace{1mm}
\noindent \textbf{Overview of Survey~2:} We created a personalized Survey~2
(instrument in Appendix~\ref{subsec:survey2})
for each participant using at most fifteen selected data elements, depending on the presence/absence of a particular class. During the survey, we first showed these data elements randomly to the participants and correspondingly asked some related questions, e.g., what are the contents of the data element and how comfortable is the participant in sharing it with people in different proxemic zones~\cite{hall1966hidden} as well as Google. Note that participants were not provided with any clue about the possibly sensitive nature of these data elements at this stage of Survey~2. Next, we gave participants a brief explanation of the respective classes from which the data elements for their personalized survey were selected. The participants also rated the accuracy of those explanations. Then, to demonstrate the possible output of automated techniques to uncover sensitive data elements, we asked participants to review a personalized Google Drive folder with named files and subfolders containing categorized possibly sensitive GVA data. Then we asked questions to measure user awareness about GVA after seeing this folder. We concluded by asking about the utility of an automated system for detecting sensitive GVA-collected data. In the end, we gave instructions to uninstall the browser extension.

\subsection{Limitations}
\noindent First, our study is limited in recruitment since we recruited only US Prolific users who primarily use Android smartphones and are familiar with GVA. In other words, we might have chosen primarily English speaking users who are also more frequent Android and GVA users than average. However, US-based users are still an important portion of the GVA user base and any privacy issue uncovered by exploring experienced GVA users possibly also affects lesser experienced users. Second, we focused on GVA users who also used an Android device as their primary smartphone. Since GVA is also available in iOS and third-party IoT devices, we might have missed those users. However, this is expected since, in this study, we aimed to investigate the most prominent users of GVA---Android users (GVA is installed by default in Android, unlike iOS). \newtext{Consequently, some of our survey participants' perceptions about data collection might not be representative of data collected by other voice assistants, which might be used in a different context (e.g., a voice assistant integrated into a children's toy} \newtext{Third, a few of our participants might consider some of our questions as probing based on both language of the question and their prior experience---introducing bias in some of our self-reported data-based results.} Lastly, our results might have underestimated privacy needs as very privacy-sensitive individuals would be unlikely to participate in a study that aimed to investigate their GVA data. Not covering such privacy-sensitive individuals is a common concern with user studies related to privacy~\cite{mondal_moving_2019}. However, we strongly feel that this work is still valuable since we unpack common privacy perceptions of GVA users regarding their data and identify possible avenues to improve data dashboards and simplify privacy controls for this data.

\section{Data Analysis}
\label{sec:analysis}

\noindent We performed both quantitative and qualitative analyses of participants' survey responses. In this section, we briefly elaborate on our data analysis process.

\vspace{1mm}
\noindent \textbf{Qualitative open coding}: We performed qualitative open coding to analyze free-text responses~\cite{qual-coding-book}. First, an author analyzed the responses to each question and created a codebook. Next, two researchers independently coded the responses using this shared codebook. Across all questions,
Cohen's kappa (inter-rater agreement) ranged from 0.769 to 1.0 signifying near-perfect agreement. At last, the coders met to resolve disagreements and finalized a code for each response.

\vspace{1mm}
\noindent \textbf{Quantitative statistical analysis}: To gain more insight into the collected quantitative data, we performed several statistical tests~\cite{statistics-analysis,statistics-effect-size}. \newtext{When the independent variable was categorical, and the dependent variable was numerical, we found all distributions were non-normal (using the Shapiro Wilkes test) and nearly all independent variables with more than two levels. Therefore, we decided to opt for the Kruskal Wallis test for comparing distributions in such cases.}
When both independent and dependent variables were categorical, we used either the $\chi^2$ test or Fisher's exact test (when individual cell values in the contingency table were $<$ 5)  to find significant correlations. We also used difference in proportions as a measure of effect size in our analysis. Apart from statistical tests, we used standard evaluation metrics such as accuracy, precision, and recall to test our prediction model~\cite{ml-book, recsys-book}.

\section{Results}
\label{sec:results}

\noindent In this section, we present results from our study on understanding user perceptions and privacy preferences regarding GVA data. \newtext{Unless otherwise specified, results in this section will correspond to self-reported data and not actual usage data.} \newtext{In specific analyses (e.g., sharing comfortability) involving audio clips and transcripts from Survey 2, we sometimes discounted very few elements due to lack of user feedback.}

\subsection{Participants}

\noindent \newtext{A} total of 80 participants completed both Survey~1 and Survey~2. We start by checking the basic demographics of those participants in this section.

\vspace{2mm}
\noindent \textbf{Basic demographics:} Our participant pool had a slight gender bias---68.8\% self-identified as male, 30\% as female, and 1.2\% as non-binary. In terms of age---30\% were 18-24 years old, 31.3\% were 25-34 years old, and 26.3\% of the participants were between 35 and 44 years. Our participants self-identified themselves with several ethnicities---66.3\% reported themselves as White, 13.8\% as Asian or Pacific Islander, 8.8\% as Black or African American and 6.3\% as Hispanic or Latino. The rest had mixed ethnicity. The majority of our participants were employed---47.5\% employed full-time and only 20\% identified as students. In our sample, 53.75\% of participants had a bachelor's degree or higher, and only 30\% were associated with computer science or a related field. Overall, our participants came from a wide demographic spread.

\vspace{2mm}
\noindent \textbf{Usage of Android smartphones:} Even though we did not specifically attempt to recruit long-time Android users, 91.3\% of our participants reported using an Android smartphone for three years or more. Furthermore, 90\% of participants also mentioned using their current Google Account on Android smartphones for three years or more. We had an active Android-user sample--- 61.3\% of participants used their smartphones daily for 2 to 6 hours and 26.3\% for 6 to 10 hours and 6.3\% daily for more than 10 hours. The participants used different smartphone applications---54.5\% participants had more than 50 apps on their phones at the time of the study.
The majority of our participants were familiar with advanced Android features such as rooting, developer options, and launchers (over 70\% participants for each). In our sample, 87.5\% of participants owned devices running recent Android versions (9 or 10) manufactured by nine different manufacturers. Overall, our participants were long term Android users, well aware of advanced features, and had moderate to high daily usage.

\subsection{Characterizing GVA usage (RQ1)}

In this subsection, we present results on general usage patterns of GVA as well as the context for such usage.

\vspace{2mm}
\noindent \textbf{General usage:} 72.5\% of our participants were long-time GVA users, with 43.8\% participants using GVA for three years or more, and 28.8\% using it for two years and 17.5\% for a year.
In terms of usage frequency, 43.8\% of participants used GVA at least once a day, 30\% used it a couple of times per week, and the remaining 26.2\% of participants used it once a week to a couple of times per month. Participants used different methods to activate GVA (with some using multiple methods)---76.3\% of participants used a hotword (e.g., ``OK Google''), and 56.3\% activated GVA by touching, pressing, or holding buttons on their device.
Interestingly, 97.5\% of participants used GVA in three broad zones: home, office, and car encompassing both professional and personal lives. Additionally, 38.8\% of participants also reported using Google smart speakers. Using \newtext{actual usage data collected in part 1 of the study (as described in Section~\ref{subsec:collection})}, we found that interactions with GVA resulted in 138,874 data elements stored in Google's servers. The median participant had 837.5 data elements, signifying \newtext{non-negligible} usage of GVA. \newtext{An overview of participants' GVA data is in Table~\ref{tab:data} and year-wise statistics are in
Table~\ref{tab:yearwise_data} (Appendix~\ref{sec:additional}).
}

\begin{table}[!t]
    \small
    \centering
    \begin{tabularx}{\linewidth}{|X|S[table-format=6.0]|S[table-format=2.0]|S[table-format=3.1]|S[table-format=5.0]|}
        \hline
        \textbf{Data element} & \centering\textbf{Total} & \textbf{Min.} & \textbf{Median} & \textbf{Max.} \\
        \hline
        \# Audio w/ transcript & 83635 & 12 & 354 & 19451 \\ \hline
        \# Only transcripts & 55243 & 0 & 273.5 & 4671 \\ \hline
        \# Ambient location & 84309 & 1 & 407.5 & 12504 \\ \hline
        Total \# data elements & 138878 & 16 & 837.5 & 22073 \\ \hline
        Age of data (yrs.) & NA & 1 & 3 & 8 \\ \hline
    \end{tabularx}
    \caption{Overview of participants' GVA data.}
    \label{tab:data}
\end{table}

\vspace{2mm}
\noindent \textbf{Understanding context of GVA usage:} To understand the context for using GVA, we analyzed participant responses to the question-\textit{For what purposes do you use Google Assistant on your Android smartphone?} from Survey 1. The common reasons for using GVA were getting local information (50), communicating with others (29), resolving a query (28), playing audio and video files (27), navigating through devices (25), controlling other devices (24), entertainment such as games, jokes, etc. (16), and planning their day (14). Thus, participants used GVA for a wide number of purposes.

\vspace{2mm}
\noindent \textbf{Usage of GVA in smartphones:}
For each of the 1,027 data elements (audio and transcript) presented in Survey 2, we asked participants to choose the device that, according to them, collected each data element (GVA can run in multiple devices).
Participants reported that 494 (73.9\%) out of 668 audio clips were collected by GVA installed on smartphones, whereas smart speakers collected only 92 clips, indicating a bias towards smartphones for GVA usage. For a non-trivial 81 clips, participants either did not recall or even did not know. The results are similar for transcripts where 229 (63.78\%) out of 359 transcripts were collected by GVA on smartphones, and smart speakers collected 57 transcripts; the participants could not recall or didn't know the source for the rest.
Thus, most data elements presented in Survey 2 were collected by GVA on smartphones. We note that this bias towards GVA use on Android smartphones could be because of our inclusion criteria since we recruited users of Android, which has GVA pre-built into it. Still, our finding hints at an important domain of heavy data collection by GVA on smartphones in a wide variety of contexts.

\vspace{2mm}
\noindent \textbf{Summary:} 73.8\% of our participants used GVA frequently (at least a couple of times per week or more). The majority of GVA usage happened in smartphones in multiple contexts, and our median participant contributed a total of 837.5 data elements. The median age of GVA data was 3 years.

\subsection{User Perceptions of GVA Data (RQ2)}

Next, we check whether participants understood how Google handled any GVA-related data. Specifically, we investigate user perceptions regarding GVA data collection and storage using data from Survey 1.

\begin{table}[!t]
    \small
    \centering
    \begin{tabularx}{\linewidth}{|l|X|}
        \hline
        \textbf{Data Type} & \textbf{Description} \\
        \hline
        audio & Audio clip of conversation\\\hline
        transcript & Transcript of conversation\\ \hline
        location & Ambient location at the time of use\\ \hline\hline
        date & Date of conversation\\ \hline
        epoch & Time of conversation\\ \hline
        noti & Notifications sent by GVA\\ \hline
        trig & Activation method (w/ or w/o hotword)\\ \hline

        \hline
    \end{tabularx}
    \caption{Types of data collected by GVA}
    \label{tab:types}
\end{table}

\vspace{2mm}
\noindent \textbf{Perceptions of overall data collection:} First, we identified (using the Google account of the authors) that there are seven different types of data collected (\newtext{at} most) by GVA. We verified these types in our automated data collected phase too. Table~\ref{tab:types} shows the seven different types of data collected (at most) by GVA. The top three are the most obviously sensitive data types (audio, transcript, location), whereas the rest can be considered metadata. Recall that we referred to these three obviously sensitive data types as data elements in this work. To understand the awareness about GVA data collection, we asked participants-\textit{Do you think that Google Assistant on your Android smartphone collects any kind of data while you are using it?} in Survey~1. 78.8\% of our participants responded affirmatively with ``Yes'', and an additional 20\% responded ``Maybe'', signifying the participants are well-aware of possible data collection by GVA.

\vspace{1mm}
\noindent \textbf{Perceptions of specific data collection:} However, then we dug deeper and asked -\textit{What pieces of data do you think are collected when you use Google Assistant on your Android smartphone?}, showing participants the list given in Table~\ref{tab:types}. Most participants expressed that GVA collected data such as the date (89.1\%) and time (85.7\%) of conversations as well as the activation method (81.5\%). Interestingly, comparatively fewer participants were aware that GVA collected sensitive data types such as the transcripts of conversations (73.9\%) and the ambient location (70.58\%). Just 61.3\% of participants believed that GVA collected audio clips of conversations, implying that a non-trivial 38.7\% of users were unaware about the collection of audio clips by GVA.

\begin{figure}[!t]
\centering
\includegraphics[width=\linewidth]{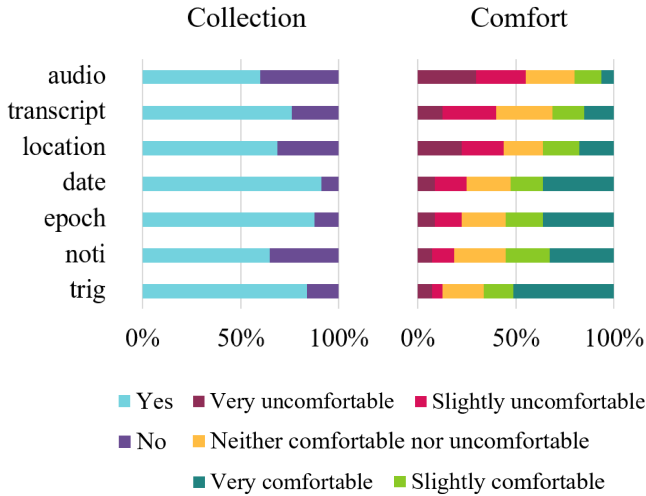}
\caption{Participant awareness of collected data types and comfort in sharing them with Google. Data types perceived to be collected/not-collected correlated with participant comfort in sharing with Google (Fisher's exact, $p < 0.0001$).
}
\label{fig:sharing}
\end{figure}

\vspace{1mm}
\noindent \textbf{Correlation between awareness of data collection and comfort in sharing data with Google:} Next, we asked participants to indicate how comfortable they would feel if GVA collected data from each of the seven data types.
The top three data types where most participants felt most uncomfortable to share with Google were audio clips of conversations (58.8\%), transcripts of conversations (45.4\%), and ambient location (45.4\%). The top three data types where most participants felt comfortable with data collection were activation method (66.4\%), notifications (56.3\%), and time of conversations (48.7\%).
Next, to check if the data types that most participants felt uncomfortable with being collected were also the ones participants were least aware of, we performed a correlation analysis. Figure~\ref{fig:sharing} presents the analysis result---participant awareness about the collection of each data type and their comfort level in sharing the data type positively correlated. These results signify that the data types for which people were less aware of collection (e.g., audio clips), people were also less comfortable with them being collected. This result indicates a superficial understanding of GVA data collection. We surmise that this shortcoming might cause a decreasing interest in GVA users to delete the GVA collected data via existing privacy controls---e.g., deleting or even browsing their stored data through the data dashboard.

\vspace{1mm}
\noindent \textbf{Perceptions of data storage:}
We asked our participants \textit{Where do you think the data, if collected by Google Assistant on your Android smartphone (and voice-enabled Google smart speakers) is stored?}\newtext{} 86.3\% of our participants correctly responded that the data is stored on Google data storage facilities (servers). However, 10\% of participants responded that the data is stored only on the respective device, whereas 3.7\% of participants responded that the data is stored completely or partially in both places. So, the majority of participants had a clear idea of about data storage practices of GVA.

\vspace{2mm}
\noindent \textbf{Summary:} Most (78.8\%) participants thought that Google collected some data using GVA, and the majority were aware of where this data is stored. However, their awareness about the type of data stored was lacking---a non-trivial fraction was unaware of the collection of sensitive data types. In fact, the participants were uncomfortable sharing the data elements they were not aware GVA was collecting (e.g., audio clips).

\subsection{Unpacking Preferred Access Control for Google to Collect GVA Data (RQ2)}

\noindent Most participants were aware that Google collects and stores some data using GVA in their servers. Thus, we investigated the desired access control rules for Google in the context of specific classes of GVA data elements.
We analyzed participant responses to this question in Survey~2 for specific data elements---\textit{After going through the audio clip/Google Assistant command, how comfortable would you feel if someone in your intimate/private/social/public relations/Google heard it/came to know about it?}.
This question checked the sharing (i.e., access control) preferences for GVA data with people in four proxemic zones--intimate, private, social public~\cite{hall1966hidden} as well as Google. Then, we used statistical analysis to check the proxemic zone closest to Google in terms of these sharing preferences. A Fisher's exact test found that there was a statistically significant correlation between desired access rules for Google and all proxemic zones (Fischer's exact $p<0.05$) across all classes of data elements from Table~\ref{tab:classes}. Then we used \textit{difference in proportions} as a measure of effect size on $2\times2$ contingency tables containing comfort data elements between a proxemic zone and google (one table for each class of data element)~\cite{statistics-effect-size}.
For each class, the zone(s) with the smallest effect size had the closest sharing preference with Google. The average effect size for each proxemic zone across all classes revealed that participants associated Google most closely with the public zone (average effect size 0.81) and farthest from the private zone (average effect size 0.92).
Table~\ref{tab:effect} (Appendix~\ref{sec:additional})
contains all the effect sizes.

\vspace{2mm}
\noindent \textbf{Summary:} Across all specific data elements, our participants expressed that the access control rules for Google should be similar to a public entity. To understand how this observation translates to actual user behaviour, we now analyze user privacy preferences for sharing specific files with Google.

\subsection{Desired Privacy Preferences of GVA Data (RQ3)}
\label{sec:corelate_class}
For specific data elements across different classes presented in Survey~2, we asked participants if they are comfortable sharing specific data elements with Google today.

\vspace{1mm}
\noindent \textbf{Users want to restrict access of Google for specific GVA data}: Our participants were uncomfortable sharing 121 (18.1\%) out of 669 audio clips and 61 (17\%) out of 358 transcripts \newtext{(presented to them in Survey 2)} with Google. These numbers indicate that participants felt uncomfortable sharing a non-trivial fraction of their GVA-collected data. Next, we will check the correlation between this preference with the medium of data collection and class of data.

\begin{figure}[!t]
\centering
\includegraphics[width=\linewidth]{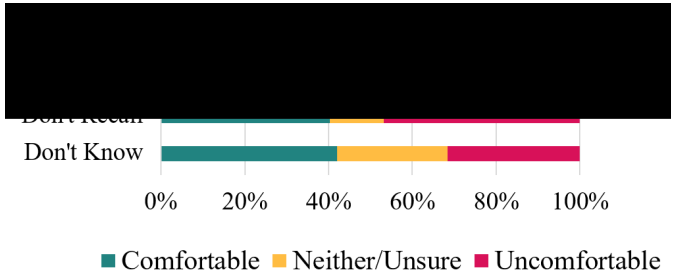}
\caption{User preferences for sharing audio clips collected by different devices with Google. Preferences for transcripts followed a similar trend.}
\label{fig:device_comfort}
\end{figure}

\vspace{2mm}
\noindent \textbf{Correlation with medium of data collection:} We checked the correlation between the device of data collection with user comfort to share data. Figure~\ref{fig:device_comfort} presents this result. Our statistical test did not reveal any significant difference in comfort across data collected via GVA on phones or smart speakers.
However, we did find a significant correlation (\textit{p} $=$ 0.00) between user knowledge of the medium of data collection and user comfort in sharing the data element (both audio and transcript) with Google. Participants felt significantly more comfortable sharing data elements where they knew about or could recall the origin of the collected data element.
Interestingly, today, Google My Activity Dashboard only shows whether a data element was collected by a Google smart speaker, completely ignoring smartphone devices. Future dashboard designs could add these missing details to make users more comfortable while viewing their data.

\vspace{2mm}
\noindent \textbf{Correlation with auto-detected classes of data element:} Participants felt most uncomfortable sharing audio clips containing regret words (24.4\% of all elements from that class), followed by audio clips having multiple speakers (21.1\%) and transcripts containing regret words (20.8\%). However, a Kruskal Wallis test revealed no significant differences between privacy preferences for data elements belonging to different classes. This result implies that perhaps these simple classes (often based on word matching) were insufficient to accurately identify the GVA-collected data elements where participants wanted to restrict access.

\vspace{1mm}
\noindent \textbf{Summary:} Participants want to restrict access for a non-trivial fraction of their GVA-collected data, which underlines a need to simplify data dashboards for identifying such data elements. Interestingly, showing users the origin of their collected data made them more comfortable sharing their data. Our simple NLP and signal processing based classes were unable to capture sensitive data elements. Thus, we need more complex tools to identify such sensitive data elements that users are uncomfortable sharing.

\subsection{Understanding Utility of Data Dashboard for privacy control (RQ4)}

\newtext{As highlighted in the previous subsection, GVA users wanted to restrict access to 17.7\% of GVA-collected data.}
Today the \textit{My Activity Dashboard} is the only Google-provided way (aside from legal interventions) for the users to delete this data (albeit post-facto) and control privacy. \newtext{Therefore, we sought to understand user perceptions regarding the
utility of this dashboard.}

\vspace{1mm}
\noindent \textbf{Perceptions of data accessibility:} First, we checked whether participants were aware that they could access the data collected by GVA. 61.3\% participants believed that the data could be accessed, while 32.5\% responded that it might be accessed, indicating the majority are at least aware of the possibility of a tool like the data dashboard.

\vspace{2mm}
\noindent \textbf{Popularity and usage of data dashboard:} Now, we check if participants knew about the My Activity dashboard.
40 (50\%) participants responded that they had heard of it, 10\% were not sure, whereas a surprisingly high 40\% of participants had never heard of the data dashboard.
Among the 40 participants who had heard of the My Activity dashboard, 4 participants had never visited it, and 33 of the remaining 36 participants visited their dashboards less than once per month. In effect, only 3 (3.8\%) out of 80 participants visited their dashboards more than once a month. According to the 36 participants who had visited their respective dashboards before our study, the top reasons for visiting it were--(i) To simply check it out (30), (ii) To view collected data (18), (iii) To change activity settings (11), and (iv) To delete some data (9). We asked 50 participants who were unsure/unaware to visit the data dashboard to check their GVA-collected data before continuing with the study.

\vspace{1mm}
\noindent \textbf{Unpacking perceptions of data dashboard:} Since all participants had explored their dashboards at least once by this point in the study, we asked them how comfortable they felt while viewing the data on their dashboard and why. On a scale of 1 to 5 (1 being very uncomfortable), the average comfortability rating was 3.26 ($\sigma$~=~1.06, median~=~3), indicating that most participants felt neither comfortable nor uncomfortable viewing the data. In fact, our qualitative analysis revealed that participants had mixed reactions to GVA data presented in the dashboard.
37.5\% of our participants were either bothered or surprised by the information collected. For instance, P31 said, ``\textit{I know that google is collecting information, but I am not 100\% comfortable to see the amount they collect. There is really no privacy.}'' 15\% of participants were glad that the data was available, and 6.25\% of participants were unsure of their choice. The remaining 41.25\% of participants were neither bothered nor surprised. For example, P12 told, ``\textit{I already know Google was collecting all of the information I saw.}'' Using a Kruskal Wallis test ($p = 0.029$), we found that participants who had heard of the Google My Activity dashboard before the study were more comfortable (N~=~40, $\mu$~=~3.525) in viewing collected data, as compared to participants who had never heard of the dashboard (N~=~32, $\mu$~=~2.875). Therefore, participants grew more comfortable with the dashboard as they became more familiar with it. Our results strongly support the need for data dashboards since participants feel more in control (and thus comfortable) when they can see and manage their data.

\vspace{1mm}
\noindent \textbf{Understanding usability of data dashboard:} To check the usability of the data dashboard, we asked our participants the question-\textit{How easy was it to reach and find what you were looking for?}, using a 5-point scale (1 being very difficult and 5 being very easy). The average rating was 4.025 ($\sigma$~=~0.899, median~=~4). Thus, most participants found My Activity Dashboard easy to reach. To get a better idea of any difficulties faced by participants during navigation, we then asked them the question-\textit{Did you face any difficulties or problems in navigating through your Google My Activity Dashboard?} Qualitative analysis showed that 8 (10\%) out of 80 participants found it hard to navigate through the dashboard to find their data. For example, P62 said, ``\textit{There is so much data it is a little overwhelming.}'' 2 participants did not check their dashboards, and 1 participant faced some problems with navigation but did not elaborate on it. The remaining 69 participants did not report any difficulties with navigation. 16.25\% of participants said they would like some assistance in using the dashboard, and another 10\% told that they might want some assistance. The remaining 73.75\% of participants indicated that they would not like any assistance. We found a positive correlation (Fisher's exact test, $p=0.048$) between the duration of using GVA (less or more than around 2 years) with the need for assistance in using the dashboard, highlighting a possible cognitive overload for long-time GVA users.

\vspace{2mm}
\noindent \textbf{Summary:} We found an interesting knowledge gap within our participants---93.8\% of participants thought their GVA-collected data can or may be accessed. However, only 50\% of the participants were aware of the data dashboard, showing a lack of actionable knowledge. Even the people who knew about data dashboards, just 3.75\% visited it regularly. In fact, \newtext{more than one-third participants
(37.5\%)} felt bothered or surprised while viewing the collected data. Quite assuringly, most participants found the dashboard easy to use; however, 10\% of participants also found it difficult to access their data. We observed that more long-time GVA users expressed a need for assistance in using the dashboard, suggesting the more they become familiar with the dashboard, the more overwhelmed they might get with the huge data collected by GVA over time.

\subsection{Moving towards Improving Utility of Data Dashboards (RQ4)}
\label{subsec:need}

Currently, Google's My Activity Dashboard provides two ways to delete collected data- (i) users can either inspect and delete each data element individually, or (ii) delete all data elements stored within a specified date-time range. The former is particularly not useful from a privacy perspective because inspecting a large number of collected data elements (most of which are non-sensitive) is quite time-consuming and laborious to be practically feasible, as seen in the previous subsection. On the other hand, the latter can help enforce privacy but is not a good option for users who might want to retain some/all of their collected data for future reference, assisting product development, etc. To assist users with finding their possibly sensitive data collected by GVA, we did a simple proof-of-concept test---around the end of Survey~2, we showed them a personalized Google drive with data elements divided into subfolders according to their auto-detected classes from Table~\ref{tab:classes} (with classes as subfolder names) and asked if a system which can show such classifications will be useful. Recall that these classes were constructed with privacy-violating scenarios in mind, and hence some of the data elements were expected to be sensitive.

\vspace{1mm}
\noindent \textbf{Recommending elements in data dashboards:} After participating in the study, 56.3\% of participants reported that they were very likely to delete some of their data collected by Google. 65\% of participants said that our classifier provided valuable assistance in finding sensitive data on Google servers, and 72.5\% told that they would recommend others to try it out if made publicly available. This percentage was 50\% higher than the 27.5\% \newtext{of} participants who expressed a need for assistance in finding sensitive data on Google servers in Survey 1, indicating a strong demand for such a recommendation system in data dashboards.

\vspace{1mm}
\noindent \textbf{The efficacy and challenge in providing recommendations:} The primary challenge that we faced while developing our sensitive content detection system was related to the accuracy of the system in assigning classes to the data elements. On a 5-point Likert scale (1 is very inaccurate and 5 is very accurate), the average rating provided by participants to our classifier was just 2.67 ($\sigma$~=~0.96, median~=~3), suggesting that most participants did not find it highly accurate. Additionally,
20 participants provided qualitative feedback regarding the study. 9 out of these 20 participants pointed out that the system accuracy could be improved. \newtext{For example, P60 stated: ``\textit{Overall I found the sensitive content system not to be very accurate. It some cases it was correct, but in more cases it was rather incorrect.}'' Despite the low perceived accuracy of the classification, we found it surprising that 65\% of participants found it useful to find sensitive content. P71 further explained the connection between classification accuracy and helpfulness of our recommendation system: ``\textit{Perhaps try improving the accuracy. I noticed that while it did get some things right, it'd periodically get things wrong. I'm not expecting the system to be perfect though but if you can improve the accuracy at all that'd be great.''}}

To better understand this result, we looked at participant accuracy scores and sharing preferences for individual data elements presented in Survey 2. \newtext{Out of 52 (65\%) participants who believed that our system provided valuable assistance in finding sensitive data (i.e. who liked our \textit{classification presentation} potentially irrespective of accuracy), we focused on 36 participants who rated at least one encountered data element~$>$=~3 for accuracy and also felt neutral or uncomfortable sharing it with Google. We observed that 29 (80.6\%) of these 36 participants found our system to be helpful (i.e. they found our classification accurate)}. So, our results hinted that even when our classifier was able to detect at least one possibly sensitive file, most participants found it useful, signifying a need for such recommendations.

\vspace{1mm}
\noindent \textbf{Summary:} Accurate recommendations of possibly sensitive data elements help users restrict access and protect the privacy of their VA-collected data. Encouragingly, recommending users to revisit even one accurately sensitive data element collected by GVA made them highly (80.6\%) likely to control their collected data, highlighting the demand for accurate, usable dashboards. This finding strongly underlines the efficacy of a highly accurate sensitive-element recommendation system to improve the utility of data dashboards. In the next section, we present the feasibility of building such an automated, highly accurate, sensitive content detection system.

\section{Feasibility of Accurately Recommending Sensitive Data in Data Dashboards}
\label{sec:recommendation}

\noindent Earlier, we saw that voice assistants like GVA collect and store large amounts of user data.
\newtext{While deleting all collected data in bulk can help avoid privacy violations, in Survey~2, participants mentioned not deleting the shown data elements for 64.8\% (674 out of 1040 data elements shown) of cases. Our open coding of their explanations revealed interesting themes---the prominent reasons for not immediately deleting these 674 data elements was that these data elements were non-sensitive (24.0\% of data elements), improving Google Assistant or Google services in general (8.2\%). For eg., P48 said, ``\textit{I don't mind Google having access to clips like this to improve their services.}'' Other reasons for not deleting collected data included having the choice to view or delete previously collected data at will (1.4\%) and personalized recommendations from Google (0.4\%). In the case of better personalized recommendations, P32 explained, ``\textit{I don't mind if Google knows hat music I listen to, especially if it improves it's music suggestion service.}'' For 17.7\% of data elements, participants did not mention any specific reason, but they wanted Google to carry out their processing and delete this data within a time frame (e.g., after 3 months).}

Even though companies provide data dashboards for users to access this data (e.g., My Activity Dashboard by Google), current dashboard designs do not offer
mechanisms for users to efficiently sift through and restrict access to specific data elements. \newtext{To that end, our results (section~\ref{subsec:need}) hint at a need for an improved human-in-the-loop (HITL) GVA data dashboard design---we envision an interface that can prioritize potentially sensitive content in the dashboard interface and assist users in controlling the privacy of their GVA-collected data.}
However, auto-detecting and restricting access to sensitive data elements to help users is also challenging as sensitivity can depend on external factors (e.g., user's age, frequency of use, other personal reasons, etc.) aside from the content of data elements. To that end, we explored the feasibility of \textit{recommending sensitive data elements} in data dashboards in a HITL scenario where the recommendation is only to help users find such data elements and not to take away their control.
Companies could leverage such recommendations to improve their data dashboards by presenting possibly sensitive data to their users for review. We envision that such data elements can be presented
either by creating a separate review section in a dashboard or changing the default ranking of shown content.

\subsection{Modelling Sensitive Content Detection as a Supervised Prediction Task}
\label{lab:predict}

\noindent Our prediction task involved predicting whether a user will perceive a particular data element collected by GVA as sensitive. For classification, our training dataset consisted of tuples ($X_i$, $Y_i$), where $X_i$ represents the feature vector, and $Y_i$ represents the sensitivity label of a data element $i$. The prediction task involved binary classification, where $Y_i$~=~1 corresponded to the `Yes' label (sensitive class) and $Y_i$~=~0 corresponded to the `No' label (not sensitive class). The feature vector $X_i$ included audio-based features, text-based features, and user-based features, all of which were captured through user survey responses and shared GVA data. The audio-based features that we used were Mel-Frequency Cepstral Coefficients (MFCC)~\cite{mfcc}, spectral contrast, tempo, and SoundNet-based features~\cite{aytar2016soundnet}. The text-based features included LIWC-based features, sentence embedding, presence of swear words, presence of regret words, sentiment-based features, emotion-based features, and presence of top 100 unigrams and bigrams. The user-based features consisted of age range and gender of users, age of Google Account, frequency and span of GVA usage, and association with computer science or a related field. The survey responses were included either as one-hot encoding or binary indicators for multiple-choice answers. A detailed description of all features is in
Appendix~\ref{subsec:features}.

To perform this classification, we explored several established supervised ML algorithms such as Support Vector Machines (SVMs), Logistic Regression (LR), Random Forest (RF), Multi-layer Perceptron (MLP), each from the scikit-learn library~\cite{pedregosa2011scikit}, along with XGBoost (XGB)~\cite{chen2016xgboost}. We compared the performance of these classifiers with two baselines. The first one was a random classifier that randomly assigned a label to each data element, where prediction probabilities for labels were chosen based on their prevalence in our dataset. For our second baseline, we used the preliminary categorization (Table~\ref{tab:classes}) of each data element as the input feature to train another XGBoost model called XGB-Class.
All model hyperparameters were optimized using grid search with 10-fold cross-validation. We found the XGB model to perform the best and thus use it to report our final performance metrics.

\subsection{Our Dataset}
\label{sec:data_ml}

Our dataset consisted of 542 audio clips and 412 transcripts.
Each data element was associated with one of three sensitivity labels by the users during Survey 2- `Yes' (sensitive class), `No' (not sensitive class), and `I am not sure' (ambiguous). \newtext{Since data sensitivity is subjective, we considered these user-assigned labels as accurate ground truth for our predictions.} The distribution of labels in the dataset was as follows- 18.34\% `Yes,' 63.94\%  `No,' and  17.72\% `I am not sure.' We were specifically interested in sensitive data elements (labeled `Yes') for our prediction task. Since the neutral label, `I am not sure,' constituted a \newtext{non-trivial fraction} of our dataset, we performed four different experiments, treating it as a separate label each time. In each experiment, we trained our best performing XGB classifier using the `I am not sure' label as a proxy for one of these four labels- `Yes' (sensitive class), `No' (not sensitive class), `Removed' (do not consider in training process), and `I am not sure' (treat as a separate class). The first three experiments were binary classification problems, whereas the last one was a three-class classification problem. \newtext{The best results were obtained associating the `I am not sure' label with the not sensitive class ('No' label), which semantically also implies a conservative prediction---recommending only data element where the classifier is certain that its sensitive. Hence, we report the results of only this experiment in the paper. The rest of the results are in
Appendix~\ref{sec:maybevariations}.
}

\subsection{Experimental Setup}
\label{sec:exp_setup}

\noindent We performed 10-fold cross-validation and reported macro-averaged precision, recall, and F1 scores for each classifier. Here, precision is defined as the ratio, TP/(TP+FP), where TP refers to the number of true positive predictions, and FP refers to the number of false positive predictions. Similarly, recall is defined as TP/(TP+FN), where FN refers to the number of false negative predictions. Since our dataset was highly skewed away from the class of our interest, we used the Synthetic Minority Oversampling TEchnique (SMOTE)~\cite{chawla2002smote} to balance our dataset before training the models. SMOTE aims to balance imbalanced datasets by oversampling or randomly replicating samples from the minority class. We used the implementation of SMOTE provided by the imbalanced-learn~\cite{lemaitre2017imbalanced} Python library.

Next, we plotted precision-recall (PR) curves (averaged over 10-folds) for each classifier to analyze the trade-off between showing a larger number of sensitive data elements to the users and the accuracy in finding such elements, reflected by recall and precision values, respectively. Maximizing both precision and recall is often not mathematically possible. In practice, a classifier with high precision and low recall returns fewer but relevant results, whereas a classifier with high recall and low precision returns many but irrelevant results. Therefore, achieving a balance between precision and recall is crucial for such classifiers to recommend as many sensitive data elements as possible to the user. A valuable heuristic to capture this trade-off between precision and recall is precision-recall area under curve (PR-AUC). A higher value of PR-AUC for a classifier shows its ability to achieve both good precision and recall. We calculated the PR-AUC values for different classifiers from their PR curves and used it as a metric to quantify their overall performance.

Finally, we used precision@k to assess different classifiers from a recommendation system's perspective.
In a practical scenario, it is unlikely that a user will go over all suggestions of sensitive data elements presented on their data dashboard. In such cases, a classifier must sort their recommendations and
minimize the number of false positives within top k recommendations. We report this value using precision@k.
Precision@k is simply the proportion of correct classifications within top k recommendations. A higher value of precision@k signifies good quality of recommendations.

\subsection{Results}

\noindent Our ML models tried to predict whether a particular data element will be perceived by the user as sensitive or not. Table~\ref{tab:result} shows the macro-averaged precision, recall, and F1 scores for all models. Across all models, XGB offered the best performance with an F1 score of 0.95, followed by LR and MLP, each of which achieved an F1 score of 0.91. The best performing baseline model was XGB-Class which achieved an F1 score of 0.54. Our XGB model outperformed the best-performing baseline model by approximately 76\%.

\begin{table}[!t]
\small
    \small
    \centering
    \begin{tabularx}{\linewidth}{|Y|Y|Y|Y|}
    \hline
    \bf Model & \bf Precision & \bf Recall & \bf F1 score \\ \hline
    \multicolumn{4}{|c|}{\bf Proposed Feature-based Models} \\\hline
    SVM & 0.90 & 0.89 & 0.89 \\ \hline
    LR & 0.92  & 0.91 & 0.91\\ \hline
    RF & 0.83  & 0.83 & 0.83\\ \hline
    MLP & 0.91  & 0.91 & 0.91 \\ \hline
    XGB & \textbf{0.96} & \textbf{0.95}  & \textbf{0.95} \\ \hline
    \multicolumn{4}{|c|}{\bf Baseline Models} \\ \hline
    Random & 0.51  & 0.52 & 0.44\\ \hline
    XGB-Class & 0.54 & 0.54 & 0.54\\ \hline
    \end{tabularx}
    \caption{Macro-averaged Precision, Recall, F1-score for all models. The highest values in each column are boldfaced.}
    \label{tab:result}
\end{table}

\begin{figure}[!t]
\centering
\includegraphics[width=\linewidth]{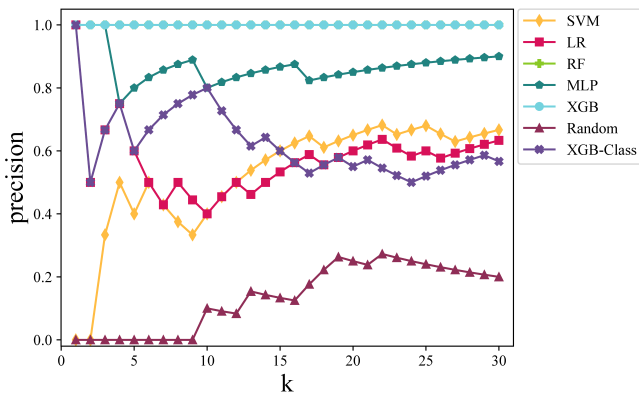}
\caption{PR curves while classifying data elements (SVM, LR, RF, MLP, and XGB are evaluated ML models, whereas Random and XGB-Class are baseline models)}
\label{fig:dm_sep}
\end{figure}

Figure~\ref{fig:dm_sep} shows the PR curves generated for all models. Once again, the XGB model performed the best, achieving a near-perfect PR-AUC value of 0.9894, followed by LR that achieved a PR-AUC value of 0.9283. The XGB model showed a significant improvement over the XGB-Class baseline model (PR-AUC~=~0.5452), outperforming it by approximately 81\%.

\begin{figure}[!t]
\centering
\includegraphics[width=0.9\linewidth]{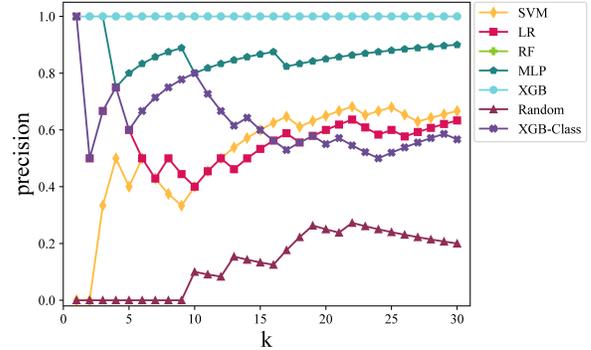}
\caption{Precision@k curves while classifying data elements (SVM, LR, RF, MLP, and XGB are evaluated ML models, whereas Random and XGB-Class are baseline models)}
\label{fig:P@K}
\end{figure}

Figure~\ref{fig:P@K} shows the precision@k curves generated for all models. Looking at the top 30 predictions, the XGB and RF models performed the best, achieving a perfect precision@30 value of 1. They were followed by MLP, which achieved a precision@30 value of 0.9. Other models such as SVM and LR had relatively poor precision@30 values comparable to the precision@30 value of 0.57 for the XGB-Class baseline model. To distinguish between the performance of XGB and RF models, we looked at their precision@k values for large values of k.
We observed a slight drop in performance while varying k from 1 to 500 for the RF model (precision@500~=~0.954), whereas the XGB model retained its performance (precision@500~=~1) even for larger values of k, highlighting the stability of XGB model (Figure~\ref{fig:P@K_500} of Appendix~\ref{sec:additional}).
Despite achieving a lower F1 score and PR-AUC value than models such as LR, the RF model offered better performance in the scenario where only a few data elements should be presented to users. \newtext{Although our best-performing XGB model achieved a high F1-score of 95\%, resourceful organisations like Google might be able to further improve accuracy in real-world deployments with additional labeled data.}

\subsection{Understanding Prediction Accuracy}

\begin{table}[!t]
    \small
    \centering
    \begin{tabularx}{\linewidth}{|c|c|X|}\hline
        \textbf{Rank} & \textbf{Feature Type} & \textbf{Name} \\ \hline
        1 & User-based & Age of Google Account\\ \hline
        2 & User-based & Frequency of GVA usage\\ \hline
        3 & Text-based & Sentiment-based\\ \hline
        4 & User-based & Age range of user\\ \hline
        5 & Audio-based & SoundNet-based\\ \hline
        6 & Text-based & LIWC-based\\ \hline
        7 & Text-based & Presence of top 100 unigrams\\ \hline
        8 & Text-based & Sentence Embedding\\ \hline
        9 & User-based & Association with CS or a related field\\ \hline
        10 & Text-based & Presence of regret words\\\hline
    \end{tabularx}
    \caption{Top 10 features as decided by the XGB model in decreasing order of importance}
    \label{tab:features}
\end{table}

Finally we analyzed the features that played the most important role in our prediction task.

\vspace{1mm}

\noindent \textbf{Many of the important features are user-based:} Table~\ref{tab:features} shows the top ten features identified by our best performing XGB classifier, in decreasing order of importance. Three out of the top five features were user-based, which highlights that user details are crucial in predicting the perceived sensitivity of data elements. Five out of the top ten features were text-based, implying that the text content of data elements is also central to the prediction task. This is in contrast with the result in Section~\ref{sec:corelate_class}, where we did not find significant differences in user privacy preferences across simple lexicon-based classes. We believe this contrast \newtext{is} because of using more involved textual features (e.g., LIWC, sentence embedding, sentiment).

\section{Concluding Discussion}\label{sec:conclusion}

\noindent In this work, we present the first study on understanding users' privacy attitudes and preferences regarding data collection by GVA.
Specifically, using a real-world data-driven approach, we unpacked users' knowledge of the data collection practices of GVA. \newtext{Previous work~\cite{farke2021dashboards} has looked into general user perceptions and reactions towards the Google My Activity data dashboard. However, we, in contrast, focused on using real-world GVA-collected data elements to elicit specific user responses}. We seek to understand whether such data dashboards \newtext{actually provide utility in controlling data} privacy through an 80-participant user study grounded into actual GVA-collected data.
Recent studies have paid increased attention to voice assistants on smart home speaker devices. Given the pervasiveness of smartphones, our results show that smartphone voice assistants can collect data in a variety of scenarios different from stationary smart speaker devices. Thus, our work sheds light on the data-centric ecosystem of voice assistants, with GVA as our test case. \newtext{Furthermore, in spite of using GVA data dashboards as our test case, many of our findings on assessing the efficacy of data dashboards and  improving their usability are generalizable to dashboards of other voice assistants.}

\vspace{1mm}
\noindent \textbf{A new direction towards usable data dashboards:} Our results establish a definite need for better data dashboards while acknowledging the utility of the current one. As a first step, our results hint at the fact that users have just superficial knowledge about the data collection and storage practices of GVA. Although data dashboards help to raise awareness about the total collected data by GVA, the huge amount of data does not help. Long-term users would need automated assistance to review more sensitive data elements. Thus, our results underline a need to make these data dashboards more usable by helping users uncover sensitive data elements. Our user feedback and accurate classification results identify that machine-learning based human-in-the-loop systems might significantly help the cause. \newtext{To that end, we identified the top ten most important features for this prediction task. In addition to text-based and audio-based features already available to the VA platform, our results highlight that user-based features can also play an important role in identifying sensitive content. We believe that a handful of these user-based features unavailable to the VA platform could be collected by directly asking the users as part of an initial setup process (while mentioning this will assist the users in finding their sensitive data elements).} In fact, our second survey, which aimed to raise awareness about different types of data collected by GVA and stored by Google increased user awareness about GVA collected data.

However, there is much left to explore in this direction, including the presentation of these recommendations to the users and checking the efficacy of interface nudges using these recommendations. \newtext{For instance, our HITL design to improve usability focused on assisting users in uncovering potentially sensitive elements. A potential future work is creating and evaluating a query system in parallel to this recommender system. Such a system could assist users in sifting through the GVA-collected data efficiently and further improve the usability of data dashboards.} Thus, we strongly feel our work paves the way to build more usable data dashboards for better assisting users and takes a step forward to bringing transparency to the data ecosystem of voice assistants.

\section{Acknowledgments}

\noindent \newtext{We thank the anonymous reviewers and our shepherd Camille Cobb for their valuable feedback. We also thank Shalmoli Ghosh for her help with an earlier iteration of this work and Niloy Ganguly for the discussion early in the project. The experiments in this work were funded by Huawei Technologies India Private Limited via the ADUL project.}

\bibliographystyle{plain}
\bibliography{one-file-audiosense-final-extended}

\appendix
\section{Instruments}
This section contains the survey instruments that we used during different parts of the study. \textcolor{gray}{[Grayed-out elements within square brackets]} present additional information about a section, question, or text prompt.

\subsection{Pre-screening Instrument}
\label{subsec:pre-screening}

\begin{scriptsize}
\textbf{Consent form}

\noindent You will be asked a few general questions in this study mainly regarding your usage of Android smartphones and Google Assistant. This is a survey to find eligible participants for our primary research study (that will be deployed shortly on Prolific) titled ``Participate in a research study and learn what data Google Assistant has stored about you''. Your Prolific ID will be used to distribute payment to you, and if eligible, to allow you to participate in our primary research study. It will not be used for any other purposes. No sensitive data will be collected. Any reports and presentations about the findings from this study will not include your name or any other information that could identify you. We may share the data collected in this study with other researchers doing future studies.

\noindent For additional questions about this research, you may contact Anonymous at anon@anon.

\noindent Please indicate, in the question below, that you have read and understood this consent form, and you agree to participate in this online research study.\\

\noindent Please indicate your consent before proceeding.
$\bigcirc$ I consent, begin the study
$\bigcirc$ I do not consent, I do not wish to participate in the study\\

\noindent \textbf{Do not consent} \textcolor{gray}{[If consent not provided]}

\noindent As you do not wish to participate in this study, please return your submission on Prolific by selecting the ``Stop without completing button''.\\

\noindent \textbf{A few general questions} \textcolor{gray}{[If consent provided]}

\noindent Please enter your Prolific ID.\_\_\_\\

\noindent What is your age?
$\bigcirc$ Younger than 18
$\bigcirc$ 18-24
$\bigcirc$ 25-34
$\bigcirc$ 35-44
$\bigcirc$ 45-54
$\bigcirc$ 55-64
$\bigcirc$ 65 or older
$\bigcirc$ I don't know
$\bigcirc$ Prefer not to say\\

\noindent Have you been using (and currently use) an Android smartphone as your primary smartphone for over one year?
$\bigcirc$ Yes
$\bigcirc$ No
$\bigcirc$ I am not sure\\

\noindent Do you currently have Google Assistant software installed on an Android smartphone? \textcolor{gray}{You can know more about Google Assistant software from the link given below- \url{https://www.digitaltrends.com/mobile/what-is-google-assistant/}}
$\bigcirc$ Yes
$\bigcirc$ No
$\bigcirc$ I am not sure\\

\noindent Do you frequently (a few times in a month or more) use Google Assistant software on an Android smartphone? \textcolor{gray}{You can know more about Google Assistant software from the link given below- \url{https://www.digitaltrends.com/mobile/what-is-google-assistant/}}
$\bigcirc$ Yes
$\bigcirc$ No
$\bigcirc$ I am not sure\\

\noindent Participation in our primary research study will help you learn what data Google Assistant has stored about you.\\

\noindent As part of our primary research study, will you be willing to install a Firefox browser extension developed by us? This extension will assist you in completing a task during the study. You can uninstall the browser extension as soon as the study is over.
$\bigcirc$ Yes
$\bigcirc$ No
$\bigcirc$ I am not sure\\

\noindent As part of our primary research study, will you be comfortable sharing the content of \textbf{ONLY} ``Assistant,'' ``Sound Search,'' and ``Voice and Audio'' folders stored in your Google Account which contain \textbf{ONLY} data collected by your Google Assistant? Note that Google Assistant \textbf{DOES NOT COLLECT} your personally identifiable sensitive information (e.g., mails, contacts, images, addresses, or any payment information). All the data that you share with us would only ever be processed by automated means (e.g., via a computer software written by us). No researcher will ever manually look at the raw data that you shared.
$\bigcirc$ Yes
$\bigcirc$ No
$\bigcirc$ I am not sure\\

\noindent \textbf{Thank You}

\noindent Thank you for participating in this study. If eligible, we will send you a message via Prolific to participate in our primary research study (that will be deployed shortly on Prolific). Participation in the primary research study will be voluntary.

\end{scriptsize}

\subsection{Survey 1 Instrument}
\label{subsec:survey1}

\begin{scriptsize}
\noindent \textbf{Consent form}

\noindent \textbf{STUDY TITLE:} Understanding perceptions regarding the use of Google Assistant

\noindent \textbf{PRINCIPAL INVESTIGATOR:} Anonymous

\noindent \textbf{STUDENT RESEARCHER:} Anonymous\\

\noindent \textbf{DESCRIPTION:} We are researchers at Anonymous doing a two-part research study to understand how users of Android smartphones feel about different aspects of Google Assistant, a popular voice assistant developed by Google and included in almost all Android smartphones. In this study, we are aiming to perform a data-driven investigation of Android users' perceptions regarding Google Assistant software data collection practices.

\noindent Note that we \textbf{DO NOT COLLECT} any personally identifiable information in this study from your Google account. \textbf{YOU ONLY NEED TO SHARE} your data collected and stored by Google Assistant software in the course of this study to help us understand your perceptions about that particular stored data. Specifically, we \textbf{DO NOT COLLECT} information such as your name, address, emails, email address, date of birth, phone number, images, videos, contacts, location data, ZIP code/postcode, links to social media accounts or posts. We fully abide by Prolific's rules of not collecting personally identifiable information. This study has been approved by our Institute Ethics Committee (IRB).\\

\noindent \textbf{STUDY TASKS:} In Part 1 of the study, you will first have to participate in a survey. This survey will involve answering general questions regarding your use of Android smartphones and Google Assistant. You will then be required to install our Firefox web browser extension (we will give the installation instructions after you complete the survey). In the last step of Part 1 of the study, you will be asked to share the content of \textbf{ONLY} ``Assistant'', ``Sound Search'' and ``Voice and Audio'' folders stored in your Google Account which contains data collected by your Google Assistant. You will have to log into your primary Google account (e.g., via \url{https://accounts.google.com/}) using the Firefox browser. Then our browser extension will assist you to automatically select the data required for our study. In this way, the browser extension will ensure that you do not share your entire Google-collected activity data, but just a part of it. Finally, you will be required to upload this data to your personal Google Drive and share the link to the data (you will be provided explicit instructions). This will ensure that the data collection is transparent and secure. This data collection will also be the last step of Part 1 of this study. After this step, you will be given the completion token to Part 1 of the study to get your compensation.

\noindent All the data collected by your Google Assistant that you share with us would only ever be processed by automated means (e.g., via a computer software written by us). No researcher will ever manually look at the raw data that you shared. Furthermore, we will provide you the results of our automated data processing in the course of Part 2 of the study.

\noindent After the automated processing of your data is completed, we will send you a message on Prolific to return for Part 2 of the study. Part 2 of the study will involve your participation in another final survey. You will be asked a series of questions regarding some of your data collected in Part 1 of the study. These will also be general questions, and the responses will be used to understand user perceptions regarding the use of Google Assistant.

\noindent Each part of the study should take approximately 35 minutes to complete.\\

\noindent \textbf{RISKS and BENEFITS:} The primary risk to your participation in this online study is unintended leakage of your private data collected by Google Assistant and shared with us, which might result in privacy violation and discomfort. We assure you that this data will be kept in a secure password-protected server in an encrypted format, and only the researchers will have access to it. The raw user data will not be accessed by any administrator at any point in the study, as our data processing step is completely automated. Some other risks to your participation in this online study are those associated with basic computer tasks, including boredom, fatigue, mild stress, or breach of confidentiality. The only benefit to you is the learning experience from participating in a research study. The benefit to society is the contribution to scientific knowledge.\\

\noindent \textbf{COMPENSATION:} You will be credited \$5.00 on Prolific for participation in Part 1 of the study. If you also complete Part 2 of the study (if eligible, we will contact you within one week), we will compensate you an additional \$7.00 as a bonus payment on Prolific. If you complete Part 1 of the study but not Part 2 of the study, you will only be paid for Part 1 of the study. Beyond this, Prolific does not allow for prorated compensation. In the event of an incomplete HIT (Human Intelligence Task), you must contact the research team, and compensation will be determined based on what was completed and at the researchers' discretion.\\

\noindent \textbf{PLEASE NOTE: This study contains a number of checks to make sure that participants are finishing the tasks honestly and completely. As long as you read the instructions and complete the tasks, your HIT will be approved. If you fail these checks, your HIT will be rejected.}\\

\noindent \textbf{CONFIDENTIALITY:} Your Prolific ID will be used to contact you for Part 2 of the study and distribute payment to you but will not be used for any other purposes. Any reports and presentations about the findings from this study will not include your name or any other information that could identify you. In some cases, you might provide personal stories or beliefs that we might quote or paraphrase as part of our research findings. If any personally identifiable information is found in such a case, it will be removed to protect your privacy. We may share the data collected in this study with other researchers doing future studies.\\

\noindent \textbf{SUBJECT'S RIGHTS:} Your participation is voluntary. You may stop participating at any time by closing the browser window or the program to withdraw from the study. Partial data will not be analyzed.\\

\noindent For additional questions about this research, you may contact Anonymous at anon@anon.

\noindent \textbf{Please indicate, in the question below, that you are at least 18 years old, have read and understood this consent form, and you agree to participate in this online research study.}\\

\noindent Please indicate your consent before proceeding.
$\bigcirc$ I consent, begin the study
$\bigcirc$ I do not consent, I do not wish to participate in the study\\

\noindent \textbf{Do not consent} \textcolor{gray}{[If consent not provided]}

\noindent As you do not wish to participate in this study, please return your submission on Prolific by selecting the ``Stop without completing button''.\\

\noindent \textbf{Introduction} \textcolor{gray}{[If consent provided]}

\noindent Dear participant,\\

\noindent Today, voice-enabled assistants in smartphones like Google Assistant, Amazon's Alexa, and Apple's Siri are becoming increasingly popular. We are a group of academic researchers conducting this study to understand how the users of Android-powered smartphones feel about different aspects of Google assistant, a popular voice assistant developed by Google and included in all Android smartphones. You can know more about Google Assistant software from here- \url{https://www.digitaltrends.com/mobile/what-is-google-assistant/}\\

\noindent The goal of our study is to raise user awareness about the usage of voice assistants in general and Google Assistant in particular. So, please read the instructions carefully, and answer all the questions judiciously. The survey will take approximately 35 minutes to complete.\\

\noindent Note that throughout the survey, we will be asking questions in the context of the Android smartphone and the Google Account you use most often, unless mentioned otherwise explicitly. Also note that we will be using the word ``sensitive'' in the context of sensitive personal information throughout the survey. Sensitive personal information refers to the personal information a person would prefer to stay private. This might include personal data revealing racial or ethnic origin, political opinions, religious or philosophical beliefs, biometric information, health-related information, financial information, sexual orientation, etc. Loss, misuse, modification, or unauthorized access to such information can adversely affect the privacy or welfare of an individual depending on the level of sensitivity and nature of the information.\\

\noindent The responses to this survey will be used only for academic research. We will not publish any personally identifiable information in any resulting report or academic publication. If you have any queries about the study please contact Anonymous at anon@anon.\\

\noindent \textbf{Device information and usage}

\noindent In the following section, we will ask you some general questions related to your usage of Android smartphones, your knowledge of the Android Operating System, and your current Android smartphone.\\

\noindent How long have you been using smartphones running Android operating system? $\bigcirc$ Around a few weeks
 $\bigcirc$ Around a few months
 $\bigcirc$ Around a year
 $\bigcirc$ Around 2 years
 $\bigcirc$ Around 3 years
 $\bigcirc$ More than 3 years \\

\noindent You needed a Google account to log into Google services on your current Android smartphone. How long have you been using that Google account on Android smartphones?
$\bigcirc$ Around a few weeks
$\bigcirc$ Around a few months
$\bigcirc$ Around a year
$\bigcirc$ Around 2 years
$\bigcirc$ Around 3 years
$\bigcirc$ More than 3 years\\

\noindent For what primary purposes do you use your Android smartphone? Select all that apply.
$\square$ Entertainment (playing games, streaming services etc.)
$\square$ Web Browsing
$\square$ Getting information (Google Maps, Bank services, Radio, Newspaper etc.)
$\square$ Personal usage (calling, email, taking photos etc.)
$\square$ Social networking (using Instagram, Facebook etc.)
$\square$ Other:\_\_\_\\

\noindent On average, within a 24‐hour period, approximately how many hours do you use your Android smartphone?
$\bigcirc$ Less than 2 hours
$\bigcirc$ 2 to 6 hours
$\bigcirc$ 6 to 10 hours
$\bigcirc$ More than 10 hours
$\bigcirc$ I am not sure\\

\noindent Which version of Android are you using on your device?
$\bigcirc$ Android 10
$\bigcirc$ Android 9 Pie
$\bigcirc$ Android 8.0 Oreo
$\bigcirc$ Android 7.0 Nougat
$\bigcirc$ Android 6.0 Marshmallow
$\bigcirc$ Android 5.0 Lollipop
$\bigcirc$ Android 4.4 KitKat
$\bigcirc$ Older than the versions of Android listed here
$\bigcirc$ Custom Android distributions (e.g., LineageOS, OmniROM, Paranoid Android, etc.)
$\bigcirc$ I am not sure\\

\noindent Do you know what rooting an Android device means?
$\bigcirc$ Yes
$\bigcirc$ No
$\bigcirc$ I am not sure\\

\noindent \textcolor{gray}{[Know about rooting]} Is your Android smartphone rooted?
$\bigcirc$ Yes
$\bigcirc$ No
$\bigcirc$ I am not sure\\

\noindent \textcolor{gray}{[Don't know about rooting]} Rooting is the process of allowing users of devices running Android Operating System to attain privileged control (known as root access) over various Android system components. Rooting gives users the ability to alter or replace system applications and settings, run specialized applications, and perform other operations that are otherwise inaccessible to a normal Android user (e.g., removing pre-installed applications that you cannot normally uninstall). However, rooting an Android device impacts device security and it might also void the device warranty.

\noindent If given a choice, would you prefer to root your Android smartphone?
$\bigcirc$ Yes
$\bigcirc$ No
$\bigcirc$ I am not sure\\

\noindent Approximately how many apps are installed on your Android smartphone?
$\bigcirc$ 1-25
$\bigcirc$ 26-50
$\bigcirc$ 51-75
$\bigcirc$ 76-100
$\bigcirc$ 101+
$\bigcirc$ I am not sure\\

\noindent From where do you mostly install applications for your Android smartphone?
$\bigcirc$ Play Store
$\bigcirc$ Other:\_\_\_\\

\noindent Do you ever install applications from unknown sources (downloaded from outside the Play Store or shared from another user) on your Android smartphone?
$\bigcirc$ Yes
$\bigcirc$ No
$\bigcirc$ I am not sure\\

\noindent Before installing an application on your Android smartphone, how often do you read the application provider's privacy policy for using the application?
$\bigcirc$ Always
$\bigcirc$ Sometimes
$\bigcirc$ Never
$\bigcirc$ I am not sure\\

\noindent Before installing an application on your Android smartphone, how often do you read the Android application's phone access permissions (such as Camera, Storage, etc.)?
$\bigcirc$ Always
$\bigcirc$ Sometimes
$\bigcirc$ Never
$\bigcirc$ I am not sure\\

\noindent Have you ever refused to install an Android application (even though you wanted the application's functionality) because of the permissions the application requested?
$\bigcirc$ Yes
$\bigcirc$ No
$\bigcirc$ I am not sure\\

\noindent Can you please briefly (a couple of words) mention what kind of data (such as Video files, Audio files, Pictures, Documents, Messages, etc.) do you consider most private on your smartphone?\_\_\_\\

\noindent Can you list the names of some applications currently installed on your Android smartphone that you feel might have access to your most private data?\_\_\_ \hfill \textcolor{gray}{[Optional]}\\

\noindent Do you use any of the following authentication mechanisms to lock the screen of your Android smartphone?
$\bigcirc$ Pattern
$\bigcirc$ Pin
$\bigcirc$ Password
$\bigcirc$ Swipe/Slide
$\bigcirc$ None
$\bigcirc$ Other:\_\_\_\\

\noindent Do you also use any of the following authentication mechanisms to conveniently unlock your Android smartphone? Select all that apply.
$\square$ Fingerprint Unlock
$\square$ Face Unlock
$\square$ Voice Unlock
$\square$ I don't use any of these mechanisms.
$\square$ Other:\_\_\_\\

\noindent Please tell us in brief (1-3 sentences) if you use any additional tools or mechanisms to protect the privacy of data residing in your Android smartphone.\_\_\_ \hfill \textcolor{gray}{[Optional]}\\

\noindent Are you currently using (or have previously used) antivirus or anti-malware applications on your Android smartphone?
$\bigcirc$ Yes
$\bigcirc$ No
$\bigcirc$ I am not sure\\

\noindent Have you ever heard of Developer options on an Android smartphone?
$\bigcirc$ Yes
$\bigcirc$ No
$\bigcirc$ I am not sure\\

\noindent Android Launchers are applications that can change the default look, feel, and functionality of your smartphone's home screen and app drawer. Did you know about Android Launchers prior to today's study?
$\bigcirc$ Yes
$\bigcirc$ No
$\bigcirc$ Maybe\\

\noindent Please choose the OEM (Original Equipment Manufacturer) of your Android smartphone. If you use more than one device, select the OEM for the device you currently use more often.
$\bigcirc$ Samsung
$\bigcirc$ Huawei
$\bigcirc$ Oppo
$\bigcirc$ Xiaomi
$\bigcirc$ Vivo
$\bigcirc$ Lenovo/Motorola
$\bigcirc$ LG
$\bigcirc$ Realme
$\bigcirc$ Sony
$\bigcirc$ Nokia
$\bigcirc$ Other:\_\_\_\\

\noindent \textbf{Background knowledge of voice assistants}

\noindent In the following section, we will ask you some general questions to learn more about your knowledge of voice assistants in general and Google Assistant in particular. We will also ask you some general questions related to your usage of Google Assistant.\\

\noindent A voice assistant is a software that can perform tasks or services for an individual based on commands or questions. Voice assistants can interpret human speech and respond via synthesized voices. Some of the popular voice assistants include Amazon Alexa, Apple Siri, Google Assistant, and Microsoft Cortana. Nowadays, Google Assistant is installed by default on almost all Android smartphones. Below we have provided some information about how you can start a conversation with Google Assistant.

\begin{enumerate}
    \item To start a conversation, hold the Home button or say ``Ok Google.'' On some Android devices, you can also say ``Hey Google.'' \textcolor{gray}{[Show image]}
    \item On some Android devices, you can also start a conversation by pressing the Microphone button on the Google Search widget. \textcolor{gray}{[Show image]}
\end{enumerate}

\noindent You can know more about Google Assistant software from the links given below-

\url{https://support.google.com/assistant/answer/7172657?co=GENIE.Platform\%3DAndroid&hl=en}

\url{https://www.digitaltrends.com/mobile/what-is-google-assistant/}

\url{https://9to5google.com/2019/07/26/google-replacing-voice-search/}\\

\noindent Which of the following best describes your level of familiarity with voice assistants such as Google Assistant on your Android smartphone?
$\bigcirc$ I have never heard of/noticed voice assistants before
$\bigcirc$ I know somewhat about voice assistants but have never used them before
$\bigcirc$ I know somewhat assistants and have used them before
$\bigcirc$ I know a lot about voice assistants but have never used them before
$\bigcirc$ I know a lot about voice assistants and have used them before
$\bigcirc$ Prefer not to answer
$\bigcirc$ Other:\_\_\_\\

\noindent Since how long have you been using Google Assistant on your Android smartphone?
$\bigcirc$ Around a few weeks
$\bigcirc$ Around a few months
$\bigcirc$ Around a year
$\bigcirc$ Around 2 years
$\bigcirc$ Around 3 years
$\bigcirc$ More than 3 years\\

\noindent In which environments do you use Google Assistant on your Android smartphone most often? Select all that apply.
$\square$ Home
$\square$ Office
$\square$ Car
$\square$ Other:\_\_\_\\

\noindent How often do you use Google Assistant on your Android smartphone?
$\bigcirc$ Multiple times a day
$\bigcirc$ Once a day
$\bigcirc$ A couple of times per week
$\bigcirc$ Once a week
$\bigcirc$ A couple of times per month
$\bigcirc$ Less than once per month\\

\noindent \textcolor{gray}{[If usage more than a couple of times per week]} What qualities of the Google Assistant encourage you to use it regularly on your Android smartphone? Select all that apply.
$\square$ You feel that it understands your query well
$\square$ You find its responses accurate
$\square$ You find at least some of its features useful
$\square$ You find it quick and efficient in completing tasks
$\square$ You find it easy to use
$\square$ Other:\_\_\_\\

\noindent \textcolor{gray}{[If usage less than a couple of times per week]} What qualities of the Google Assistant discourage you to use it regularly on your Android smartphone? Select all that apply.
$\square$ You feel that it doesn't understand your query well
$\square$ You find its responses inaccurate
$\square$ You don't find its features particularly useful
$\square$ You find it slow and inefficient in completing tasks
$\square$ You find it difficult to use
$\square$ Other:\_\_\_\\

\noindent For what purposes do you use Google Assistant on your Android smartphone? Select all that apply. \textcolor{gray}{To know more about what you can do with Google Assistant, please check- \url{https://support.google.com/assistant/answer/7172842?hl=en}}
$\square$ Getting local information
$\square$ Planning your day
$\square$ Playing audio and video files
$\square$ Other types of entertainment (such as Games, Jokes, etc.)
$\square$ Communicating with others
$\square$ Navigating through devices
$\square$ Controlling other devices
$\square$ Resolving a query
$\square$ Other:\_\_\_\\

\noindent How do you generally activate Google Assistant on your Android smartphone in order to use it? Select all that apply.
$\square$ Touching/Pressing/Holding Buttons on your device
$\square$ Using a hotword such as ``OK Google/ Hey Google''
$\square$ Other:\_\_\_\\

\noindent Do you think that Google Assistant on your Android smartphone frequently misunderstands what you speak?
$\bigcirc$ Yes
$\bigcirc$ No
$\bigcirc$ I am not sure\\

\noindent Do you recall changing any of the following default settings for Google Assistant on your Android smartphone? Select all that apply.
$\square$ Add/Modify languages for speaking to your Assistant
$\square$ Modify Assistant voice that your Assistant will use to respond to you
$\square$ Modify types of notifications you would like to receive from your Assistant
$\square$ Add/Modify Routines to do multiple things with one command
$\square$ I do not recall changing any of these settings
$\square$ Other:\_\_\_\\

\noindent Do you recall visiting the ``Explore'' section in Google Assistant on your Android smartphone to find out things that Google Assistant can do?
$\bigcirc$ Yes
$\bigcirc$ No
$\bigcirc$ I am not sure\\

\noindent \textbf{Usage of Google Assistant}

\noindent Besides using Google Assistant on your Android smartphone, do you also use voice-enabled Google smart speakers?
$\bigcirc$ Yes
$\bigcirc$ No
$\bigcirc$ Don't know what they are\\

\noindent \textcolor{gray}{[If use Google smart speakers]} For what purposes do you use voice-enabled Google smart speakers? Select all that apply. \textcolor{gray}{To know more about what you can do with voice-enabled Google smart speakers, please check: \url{https://support.google.com/googlenest/answer/7130274?hl=en-IN}}
$\square$ Getting local information
$\square$ Planning your day
$\square$ Playing audio and video files
$\square$ Other types of entertainment (such as Games, Jokes, etc.)
$\square$ Communicating with others
$\square$ Navigating through devices
$\square$ Controlling other devices
$\square$ Resolving a query
$\square$ Other:\_\_\_\\

\noindent \textcolor{gray}{[If use Google smart speakers]} Where have you placed these smart speakers? Select all that apply.
$\square$ Home
$\square$ Office
$\square$ Other:\_\_\_\\

\noindent \textcolor{gray}{[If use Google smart speakers]} How often do you use these smart speakers?
$\bigcirc$ Multiple times a day
$\bigcirc$ Once a day
$\bigcirc$ A couple of times per week
$\bigcirc$ Once a week
$\bigcirc$ A couple of times per month
$\bigcirc$ Less than once per month\\

\noindent \textcolor{gray}{[If use Google smart speakers]} What qualities of these smart speakers encourage you to use it?
$\square$ You feel that it understands your query well
$\square$ You find its responses accurate
$\square$ You find at least some of its features useful
$\square$ You find it quick and efficient in completing tasks
$\square$ You find it easy to use
$\square$ Other:\_\_\_\\

\noindent \textcolor{gray}{[If don't use Google smart speakers]} Do you currently own any voice-enabled Google smart speakers?
$\bigcirc$ Yes
$\bigcirc$ No\\

\noindent \textcolor{gray}{[If don't use Google smart speakers]} What qualities of these smart speakers discourage you to use it?
$\square$ You feel that it doesn't understand your query well
$\square$ You find its responses inaccurate
$\square$ You don't find its features particularly useful
$\square$ You find it slow and inefficient in completing tasks
$\square$ You find it difficult to use
$\square$ You find it too expensive for the value offered
$\square$ You are worried that the speakers are always listening
$\square$ Other:\_\_\_\\

\noindent \textcolor{gray}{[If don't know about Google smart speakers]} A smart speaker is a type of speaker and voice command device with an integrated virtual assistant that offers interactive actions and hands-free activation. These smart speakers enable us to do a variety of different things from controlling wireless-enabled lights to ordering various products and takeaways online, streaming music as well as providing information like the weather forecast, time, date, and many more things. Some of the most well-known smart speakers are Amazon Echo and Google Nest. In this study, we will be focusing on the usage of Google's smart speakers. \textcolor{gray}{[Show image]}\\

\noindent \textcolor{gray}{[If don't know about Google smart speakers]} Now that you know about smart speakers, would you be interested in purchasing voice-enabled Google smart speakers?
$\bigcirc$ Yes
$\bigcirc$ No
$\bigcirc$ Maybe\\

\noindent \textcolor{gray}{[If don't know about Google smart speakers]} If you were provided these smart speakers, what purposes would you use them for? Select all that apply. \textcolor{gray}{To know more about what you can do with voice-enabled Google smart speakers, please check: \url{https://support.google.com/googlenest/answer/7130274?hl=en-IN}}
$\square$ Getting local information
$\square$ Planning your day
$\square$ Playing audio and video files
$\square$ Other types of entertainment (such as Games, Jokes, etc.)
$\square$ Communicating with others
$\square$ Navigating through devices
$\square$ Controlling other devices
$\square$ Resolving a query
$\square$ Other:\_\_\_\\

\noindent Approximately how long ago did you create the primary Google Account you use with your Android smartphone?
$\bigcirc$ A few weeks
$\bigcirc$ A few months
$\bigcirc$ Around a year
$\bigcirc$ Around 2 years
$\bigcirc$ Around 3 years
$\bigcirc$ More than 3 years\\

\noindent What kind of information do you recall having shared with Google Assistant on your Android smartphone while using it? Select all that apply.
$\square$ Basic information (such as nickname, birthday \& phone number)
$\square$ Most visited places (such as Home or Work)
$\square$ Preferred mode of transportation (such as car, public transport or walk)
$\square$ Family \& other important contacts
$\square$ Payment methods \& purchase approvals
$\square$ Reservations for flights, hotels \& events
$\square$ Details of purchases made using Google Search, Maps \& Google Assistant
$\square$ I don't recall having shared any of the above mentioned information
$\square$ Other:\_\_\_\\

\noindent Do you recall being engaged in a meaningless conversation with Google Assistant on your Android smartphone?
$\bigcirc$ Yes
$\bigcirc$ No
$\bigcirc$ I am not sure\\

\noindent Do you recall talking in a funny accent with Google Assistant on your Android smartphone?
$\bigcirc$ Yes
$\bigcirc$ No
$\bigcirc$ I am not sure\\

\noindent Do you recall using inappropriate language while talking with Google Assistant on your Android smartphone?
$\bigcirc$ Yes
$\bigcirc$ No
$\bigcirc$ I am not sure\\

\noindent Do you recall using Google Assistant on your Android smartphone while also having a conversation with someone in the background?
$\bigcirc$ Yes
$\bigcirc$ No
$\bigcirc$ I am not sure\\

\noindent Do you think any other person might have used Google Assistant on your Android smartphone at some point in time?
$\bigcirc$ Yes
$\bigcirc$ No
$\bigcirc$ I am not sure\\

\noindent Can you please briefly (1-3 sentences) explain your answer choice for the previous question?\_\_\_\\

\noindent Do you recall using Google Assistant on your Android smartphone with other people such as friends, family, or colleagues talking in the background?
$\bigcirc$ Yes
$\bigcirc$ No
$\bigcirc$ I am not sure\\

\noindent Do you recall using Google Assistant on your Android smartphone in places with clearly audible background sounds, such as a factory or a club?
$\bigcirc$ Yes
$\bigcirc$ No
$\bigcirc$ I am not sure\\

\noindent Do you recall having accidentally activated Google Assistant on your Android smartphone?
$\bigcirc$ Yes
$\bigcirc$ No
$\bigcirc$ I am not sure\\

\noindent Do you think it is possible that Google Assistant on your Android smartphone might have been accidentally activated in any of the above scenarios, without you being aware of such an incident?
$\bigcirc$ Yes
$\bigcirc$ No
$\bigcirc$ I am not sure\\

\noindent Do you think that Google Assistant on your Android smartphone collects any kind of data while you are using it?
$\bigcirc$ Yes
$\bigcirc$ No
$\bigcirc$ Maybe\\

\noindent How comfortable would you feel if Google Assistant on your Android smartphone stored audio clips (e.g., voice recordings) from each of the above scenarios (listed in the leftmost column below)? \textcolor{gray}{[Matrix-style grid with the following rows]}

\noindent Answer choices for each row-
$\bigcirc$ Very comfortable
$\bigcirc$ Slightly comfortable
$\bigcirc$ Neither comfortable nor uncomfortable
$\bigcirc$ Slightly uncomfortable
$\bigcirc$ Very uncomfortable

\begin{itemize}
    \item Talking in a funny accent with Google Assistant
    \item Small talk with Google Assistant
    \item Using inappropriate language while talking Google Assistant
    \item Talking with someone in the background while using Google Assistant
    \item Other person using your Google Assistant
    \item Using Google Assistant with other people talking in the background
    \item Using Google Assistant in places with clearly audible background sounds
    \item Accidental activation of Google Assistant
\end{itemize}

\noindent Do you think that these audio clips could possibly contain any sensitive or personal information?
$\bigcirc$ Yes
$\bigcirc$ No
$\bigcirc$ Maybe\\

\noindent Can you please briefly (1-3 sentences) explain your answer choice for the previous question?\_\_\_\\

\noindent How comfortable would you feel if Google Assistant on your Android smartphone stored commands (e.g., conversation with Google Assistant converted to text) from each of the above scenarios (listed in the leftmost column below)? \textcolor{gray}{[Matrix-style grid with the following rows]}

\noindent Answer choices for each row-
$\bigcirc$ Very comfortable
$\bigcirc$ Slightly comfortable
$\bigcirc$ Neither comfortable nor uncomfortable
$\bigcirc$ Slightly uncomfortable
$\bigcirc$ Very uncomfortable

\begin{itemize}
    \item Talking in a funny accent with Google Assistant
    \item Small talk with Google Assistant
    \item Using inappropriate language while talking Google Assistant
    \item Talking with someone in the background while using Google Assistant
    \item Other person using your Google Assistant
    \item Using Google Assistant with other people talking in the background
    \item Using Google Assistant in places with clearly audible background sounds
    \item Accidental activation of Google Assistant
\end{itemize}

\noindent Do you think that these commands could possibly contain any sensitive or personal information?
$\bigcirc$ Yes
$\bigcirc$ No
$\bigcirc$ Maybe\\

\noindent Can you please briefly (1-3 sentences) explain your answer choice for the previous question?\_\_\_\\

\noindent Would you be comfortable sharing this data (from each of the scenarios discussed above) with Google?
$\bigcirc$ Yes
$\bigcirc$ No
$\bigcirc$ I am not sure \\

\noindent For each of the following cases when do you think it would be acceptable for Google to collect and store your data from Google Assistant on your Android smartphone? \textcolor{gray}{[Matrix-style grid with the following rows]}

\noindent Answer choices for each row-
$\bigcirc$ Always acceptable
$\bigcirc$ Sometimes acceptable
$\bigcirc$ Never acceptable

\begin{itemize}
    \item If you have explicitly consented Google to collect your data
    \item If Google does not notify you while collecting your data
    \item If Google explicitly notifies you while collecting some/all of your data
    \item If Google anonymizes (removes personally identifiable information from) your data before storing it on its data storage facilities
    \item If you get financial rewards (in the form of coupon discounts, gift cards, etc.) for sharing your data with Google
    \item If your data is used to improve the performance of Google Assistant software
    \item If your data is used to develop new features for Google Assistant software
    \item If your data is stored indefinitely by Google on its data storage facilities
    \item If your data is not stored indefinitely by Google on its data storage facilities
\end{itemize}

\noindent Can you please briefly (1-3 sentences) explain your answer choice for the cases where you chose ``Sometimes acceptable'' and ``Never acceptable'' in the previous question?\_\_\_ \hfill \textcolor{gray}{[Optional]}\\

\noindent \textbf{Thoughts about data storage}

\noindent In the following section, we will ask you some general questions to learn more about your knowledge of data collection practices of Google Assistant.\\

\noindent \textcolor{gray}{[If don't know that Google Assistant collects data]} At this stage of the study, we would like to inform you that by default Google Assistant on your Android smartphone (and voice-enabled Google smart speakers) collects a variety of user data, in order to improve its services.\\

\noindent What pieces of data do you think are collected when you use Google Assistant on your Android smartphone? Select all that apply.
$\square$ The conversation converted to text (transcript of the conversation)
$\square$ The conversation as complete audio clips
$\square$ Date of having conversation
$\square$ Time of having conversation
$\square$ Ambient (Approximate) location
$\square$ Details of how Google Assistant was activated (via button/speech)
$\square$ Details of notifications sent by Google Assistant
$\square$ Other:\_\_\_\\

\noindent How comfortable would you feel if Google Assistant on your Android smartphone collected data for each of the following categories? \textcolor{gray}{[Matrix-style grid with the following rows]}

\noindent Answer choices for each row-
$\bigcirc$ Very comfortable
$\bigcirc$ Slightly comfortable
$\bigcirc$ Neither comfortable nor uncomfortable
$\bigcirc$ Slightly uncomfortable
$\bigcirc$ Very uncomfortable

\begin{itemize}
    \item The conversation converted to text (transcript of the conversation)
    \item The conversation as complete audio clips
    \item Date of having conversation
    \item Time of having conversation
    \item Ambient (Approximate) location
    \item Details of how Google Assistant was activated (via button/speech)
    \item Details of notifications sent by Google Assistant
\end{itemize}

\noindent Can you please briefly (1-3 sentences) explain your answer choice for the categories where you chose ``Slightly uncomfortable'' and ``Very uncomfortable'' in the previous question?\_\_\_ \hfill \textcolor{gray}{[Optional]}\\

\noindent What pieces of data do you think are collected when you use voice-enabled Google smart speakers? Select all that apply.
$\square$ The conversation converted to text (transcript of the conversation)
$\square$ The conversation as complete audio clips
$\square$ Date of having conversation
$\square$ Time of having conversation
$\square$ Ambient (Approximate) location
$\square$ Details of how Google Assistant was activated (via button/speech)
$\square$ Details of notifications sent by Google Assistant
$\square$ Other:\_\_\_\\

\noindent How comfortable would you feel if voice-enabled Google smart speakers collected data for each of the following categories? \textcolor{gray}{[Matrix-style grid with the following rows]}

\noindent Answer choices for each row-
$\bigcirc$ Very comfortable
$\bigcirc$ Slightly comfortable
$\bigcirc$ Neither comfortable nor uncomfortable
$\bigcirc$ Slightly uncomfortable
$\bigcirc$ Very uncomfortable

\begin{itemize}
    \item The conversation converted to text (transcript of the conversation)
    \item The conversation as complete audio clips
    \item Date of having conversation
    \item Time of having conversation
    \item Ambient (Approximate) location
    \item Details of how Google Assistant was activated (via button/speech)
    \item Details of notifications sent by Google Assistant
\end{itemize}

\noindent Can you please briefly (1-3 sentences) explain your answer choice for the categories where you chose ``Slightly uncomfortable'' and ``Very uncomfortable'' in the previous question?\_\_\_ \hfill \textcolor{gray}{[Optional]}\\

\noindent Where do you think the data, if collected by Google Assistant on your Android smartphone (and voice-enabled Google smart speakers) is stored?
$\bigcirc$ On the respective device
$\bigcirc$ On Google data storage facilities
$\bigcirc$ Other:\_\_\_\\

\noindent Do you recall getting an email from Google recently (within the last week) with the subject ``We've updated your voice and audio setting''?
$\bigcirc$ Yes
$\bigcirc$ No
$\bigcirc$ I am not sure\\

\noindent \textcolor{gray}{[If recall getting this email]} Did you read this email sent by Google?
$\bigcirc$ Yes, I read this email carefully
$\bigcirc$ Yes, I quickly glanced over the email
$\bigcirc$ No, I did not read this email
$\bigcirc$ I am not sure\\

\noindent \textcolor{gray}{[If read or glanced over this email]} We have attached a screenshot of the email below for your reference. Please answer the questions that follow. \textcolor{gray}{Show screenshot}\\

\noindent \textcolor{gray}{[If read or glanced over this email]} Did you click on the ``REVIEW SETTING'' button to review the audio recordings setting, as highlighted by Google in the email?
$\bigcirc$ Yes
$\bigcirc$ No
$\bigcirc$ I am not sure\\

\noindent \textcolor{gray}{[If unsure, did not read, or do not recall getting this email]} Google is rolling out an update for the default voice and audio setting of its users. We have attached a screenshot of the email below for your reference. Please read this email carefully and answer the questions that follow. \textcolor{gray}{Show screenshot}\\

\noindent Based on the explanation provided in the email, what do you think the current audio recordings setting is for your Google Account?
$\bigcirc$ Google will save audio recordings in your Google Account
$\bigcirc$ Google will not save audio recordings in your Google Account
$\bigcirc$ I am not sure\\

\noindent Can you please briefly (1-3 sentences) explain why do you think Google updated the setting for voice and audio recordings in your Google Account?\_\_\_\\

\noindent Did you find the language of the email clear and easy to understand?
$\bigcirc$ Yes
$\bigcirc$ No
$\bigcirc$ I am not sure\\

\noindent We would now like to tell you that until recently, Google Assistant on your Android smartphone used to collect all the pieces of data mentioned on the previous page, including voice recordings of the conversation as audio clips, in order to improve its services. However, recently Google has stopped collecting these audio clips due to several privacy concerns. The email which was showed to you in an earlier section was meant to notify users about this change. The collected data is then sent to Google data storage facilities where it is used by Google to develop and improve its audio recognition technologies and the Google services that use them, like Google Assistant. Due to privacy concerns, Google is also working on new techniques such as Federated Learning to improve its services without having the need of uploading user data to its data storage facilities. Using Federated Learning, Google can perform required computations on your device and collect just the results of these computations, instead of complete user data.\\

\noindent In general, which of the following privacy policies would you prefer to manage the privacy of your data from Google Assistant on your Android smartphone  (e.g., audio clips, commands, etc.) in the future?
$\bigcirc$ Give consent to Google to collect and store all data indefinitely, until you choose to delete it
$\bigcirc$ Give consent to Google to collect and store all data, but delete all data older than 3 months
$\bigcirc$ Give consent to Google to collect and store all data, but delete all data older than 18 months
$\bigcirc$ Store all data locally on your device, perform necessary computations required by Google on device and share just the computation results instead of complete data with Google
$\bigcirc$ Do not give consent to Google to collect and store all data, and delete any data currently stored by Google immediately
$\bigcirc$ None of the above\\

\noindent Do you believe that Google ensures the security of the data they store?
$\bigcirc$ Yes
$\bigcirc$ No
$\bigcirc$ I am not sure\\

\noindent On a scale of 1 to 5, with 1 being the least and 5 being the most, how much do you trust Google that it would not use your data for malicious purposes?
$\bigcirc$ 1
$\bigcirc$ 2
$\bigcirc$ 3
$\bigcirc$ 4
$\bigcirc$ 5\\

\noindent Do you think that you can access your data collected by Google?
$\bigcirc$ Yes
$\bigcirc$ No
$\bigcirc$ Maybe\\

\noindent \textbf{Questions specific to Google Activity}

\noindent In the following section, we will ask you some general questions to learn more about your knowledge of Google-collected activity data.\\

\noindent Did you know that you can access all the data that Google collected about you including the data collected by Google Assistant?
$\bigcirc$ Yes
$\bigcirc$ No
$\bigcirc$ Maybe\\

\noindent Did you know that you can access your data collected by Google Assistant by visiting ``Settings'' in your Google Assistant Application?
$\bigcirc$ Yes
$\bigcirc$ No
$\bigcirc$ Maybe\\

\noindent Have you ever heard of Google My Activity dashboard?
$\bigcirc$ Yes
$\bigcirc$ No
$\bigcirc$ I am not sure\\

\noindent \textcolor{gray}{[If heard of Google My Activity dashboard]} How did you first come to know about Google My Activity dashboard? Select all that apply.
$\square$ Search Engine
$\square$ Advertisements
$\square$ Social Media
$\square$ Radio
$\square$ TV
$\square$ Print Media
$\square$ Word of mouth
$\square$ Other:\_\_\_\\

\noindent \textcolor{gray}{[If heard of Google My Activity dashboard]} Have you ever visited your Google My Activity dashboard?
$\bigcirc$ Yes
$\bigcirc$ No
$\bigcirc$ I am not sure\\

\noindent \textcolor{gray}{[If visited Google My Activity dashboard before]} Why did you visit your Google My activity dashboard? Select all that apply.
$\square$To view collected data
$\square$To delete some data
$\square$To change Activity settings
$\square$To simply check it out
$\square$Other:\_\_\_\\

\noindent \textcolor{gray}{[If visited Google My Activity dashboard before]} Can you please briefly (1-3 sentences) explain your answer choice for the previous question?\_\_\_\\

\noindent \textcolor{gray}{[If visited Google My Activity dashboard before]} How often do you visit your Google My Activity dashboard?
$\bigcirc$ Multiple times a day
$\bigcirc$ Once a day
$\bigcirc$ A couple of times per week
$\bigcirc$ Once a week
$\bigcirc$ A couple of times per month
$\bigcirc$ Less than once per month
$\bigcirc$ Other:\_\_\_\\

\noindent \textcolor{gray}{[If unsure/never heard of/visited Google My Activity dashboard before]} Google My Activity dashboard allows users to view and delete their collected data across all Google products and services. Please click the link below to visit your My Activity dashboard. You would have to login to your Google Account. Make sure to check out data across several product categories by using the Filter option provided. \url{https://myactivity.google.com/myactivity}\\

\noindent \textcolor{gray}{[If unsure/never heard of/visited Google My Activity dashboard before]} Among the data collected across several product categories, which ones did you check out? Select all that apply.
$\square$ Ads
$\square$ Android
$\square$ Assistant
$\square$ Books
$\square$ Chrome
$\square$ Developers
$\square$ Discover
$\square$ Drive
$\square$ Gmail
$\square$ Google Apps
$\square$ Google Cloud
$\square$ Google Lens
$\square$ Google News
$\square$ Google Pay
$\square$ Google Play Games
$\square$ Google Play Movies \& TV
$\square$ Google Play Music
$\square$ Google Play Store
$\square$ Help
$\square$ Image Search
$\square$ Maps
$\square$ News
$\square$ Search
$\square$ Shopping
$\square$ Sound Search
$\square$ Video Search
$\square$ Voice and Audio
$\square$ YouTube
$\square$ Other:\_\_\_\\

\noindent \textcolor{gray}{[If unsure/never heard of/visited Google My Activity dashboard before]} Would you like to visit your Google My Activity dashboard more often in the future?
$\bigcirc$ Yes
$\bigcirc$ No
$\bigcirc$ Maybe\\

\noindent \textcolor{gray}{[If unsure/never heard of/visited Google My Activity dashboard before]} Can you please briefly (1-3 sentences) explain your answer choice for the previous question?\_\_\_\\

\noindent How comfortable did you feel while viewing the collected data on your Google My Activity dashboard?
\textcolor{gray}{Very uncomfortable}
$\bigcirc$ 1
$\bigcirc$ 2
$\bigcirc$ 3
$\bigcirc$ 4
$\bigcirc$ 5
\textcolor{gray}{Very comfortable}\\

\noindent Can you please briefly (1-3 sentences) explain your answer choice for the previous question?\_\_\_\\

\noindent How easy was it to reach and find what you were looking for?
\textcolor{gray}{Very difficult}
$\bigcirc$ 1
$\bigcirc$ 2
$\bigcirc$ 3
$\bigcirc$ 4
$\bigcirc$ 5
\textcolor{gray}{Very easy}\\

\noindent Did you face any difficulties or problems in navigating through your Google My Activity dashboard?\_\_\_\\

\noindent Would you like some assistance in using your Google My Activity dashboard?
$\bigcirc$ Yes
$\bigcirc$ No
$\bigcirc$ Maybe\\

\noindent Would you like some assistance in downloading this data collected by Google?
$\bigcirc$ Yes
$\bigcirc$ No
$\bigcirc$ Maybe\\

\noindent Would you like some assistance in finding sensitive data and deleting it from Google data storage facilities?
$\bigcirc$ Yes
$\bigcirc$ No
$\bigcirc$ Maybe\\

\noindent Recall that we informed you about the fact that by default Google Assistant on your Android smartphone collects a variety of user data. Would you be willing to delete some of the files from Google data storage facilities?
$\bigcirc$ Yes
$\bigcirc$ No
$\bigcirc$ Maybe\\

\noindent Can you please briefly (1-3 sentences) explain your answer choice for the previous question?\_\_\_\\

\noindent Google provides users with a setting to delete all collected data older than 3 months/18 months automatically. Would you be willing to opt for this?
$\bigcirc$ Yes
$\bigcirc$ No
$\bigcirc$ Maybe\\

\noindent Did you know about this option prior to today's study?
$\bigcirc$ Yes
$\bigcirc$ No
$\bigcirc$ I am not sure\\

\noindent Do you feel that tech companies, in general, tend to make it harder for users to access their collected data?
$\bigcirc$ Yes
$\bigcirc$ No
$\bigcirc$ Maybe\\

\noindent Have you ever gone through Google's privacy policy?
$\bigcirc$ Yes
$\bigcirc$ No
$\bigcirc$ I am not sure\\

\noindent On a scale of 1 to 5, with 1 being the least and 5 being the most, how much do you trust Google that it would not use your data for malicious purposes, after participating in this study?
$\bigcirc$ 1
$\bigcirc$ 2
$\bigcirc$ 3
$\bigcirc$ 4
$\bigcirc$ 5\\

\noindent \textbf{General privacy perception related questions}

\noindent In the following section, we will ask you some general questions to learn more about your perceptions regarding data collection, disclosure, and management policies of technology companies as a whole.\\

\noindent On a scale of 1 to 5, with 1 being strongly disagree and 5 being strongly agree, express your level of agreement for each of the statements given below. Here 1 indicates ``Strongly disagree'', and 5 indicates ``Strongly agree''.
\textcolor{gray}{[Matrix-style grid with the following rows]}

\noindent Answer choices for each row-
$\bigcirc$ 1
$\bigcirc$ 2
$\bigcirc$ 3
$\bigcirc$ 4
$\bigcirc$ 5

\begin{itemize}
    \item Companies seeking information online should disclose the way the data are collected, processed, and used.
    \item A good consumer online privacy policy should have a clear and conspicuous disclosure.
    \item It is very important to me that I am aware and knowledgeable about how my personal information will be used.
    \item It usually bothers me when online companies ask me for personal information.
    \item When online companies ask me for personal information, I sometimes think twice before providing it.
    \item It bothers me to give personal information to so many online companies.
    \item I'm concerned that online companies are collecting too much personal information about me.
    \item I feel that I should have more control over how online companies collect and use my personal information.
\end{itemize}

\noindent \textbf{Demographics}

\noindent With what gender do you identify?
$\bigcirc$ Female
$\bigcirc$ Male
$\bigcirc$ Non-binary
$\bigcirc$ Prefer not to answer\\

\noindent What is your age?
$\bigcirc$ Younger than 18
$\bigcirc$ 18-24
$\bigcirc$ 25-34
$\bigcirc$ 35-44
$\bigcirc$ 45-54
$\bigcirc$ 55-64
$\bigcirc$ 65 or older
$\bigcirc$ I don't know
$\bigcirc$ Prefer not to say\\

\noindent Can you please specify your ethnicity? Select all that apply.
$\square$ White
$\square$ Hispanic or Latino
$\square$ Black or African American
$\square$ Native American or American Indian
$\square$ Asian/Pacific Islander
$\square$ Prefer not to answer
$\square$ Other:\_\_\_\\

\noindent What is the highest degree or level of school you have completed?
$\bigcirc$ No formal education
$\bigcirc$ Nursery school to 8th grade
$\bigcirc$ Some high school, no diploma
$\bigcirc$ High school graduate, diploma or the equivalent
$\bigcirc$ Some college credit, no degree
$\bigcirc$ Trade, technical, or vocational training
$\bigcirc$ Associate's degree
$\bigcirc$ Bachelor's degree
$\bigcirc$ Master's degree
$\bigcirc$ Professional degree
$\bigcirc$ Doctorate degree
$\bigcirc$ Prefer not to answer\\

\noindent What is your employment status?
$\bigcirc$ Student
$\bigcirc$ Full-time employed
$\bigcirc$ Part-time employed
$\bigcirc$ Not employed
$\bigcirc$ Retired
$\bigcirc$ Prefer not to answer\\

\noindent Are you majoring in or do you have a degree or job in computer science, computer engineering, information technology, or a related field?
$\bigcirc$ Yes
$\bigcirc$ No
$\bigcirc$ Prefer not to answer

\end{scriptsize}

\subsection{Survey 2 Instrument}
\label{subsec:survey2}

\begin{scriptsize}

\noindent \textbf{Introduction}

\noindent This survey is the final part of our two-part study on investigating the data collected by Google Assistant software. You successfully completed Part 1 earlier and shared a part of your Google My Activity data. We have now completed processing your data. The aim of this part of the survey is to understand your perceptions regarding a part of the data that you shared earlier. In this survey, we will show you a few audio clips from your Google My Activity data and ask you some questions regarding them. We will also tell you about some interesting observations we made from your data, and ask some general questions regarding the same.\\

\noindent Note that we will be using the word ``sensitive'' in the context of sensitive personal information throughout the survey. Sensitive personal information refers to the personal information a person would prefer to stay private. This might include personal data revealing racial or ethnic origin, political opinions, religious or philosophical beliefs, biometric information, health-related information, financial information, sexual orientation, etc. Loss, misuse, modification, or unauthorized access to such information can adversely affect the privacy or welfare of an individual depending on the level of sensitivity and nature of the information.\\

\noindent Please read the instructions carefully, and answer all the questions judiciously. The survey will take approximately 35 minutes to complete.\\

\noindent Please re-enter your Prolific ID.\_\_\_\\

\noindent \textbf{Questions about collected data}

\noindent We will now show you some of the data that was present in your Google-collected activity data and ask some questions related to it.\\

\noindent \textcolor{gray}{[Repeat for all audio clips presented]} The following audio clip was found in your Google My Activity data. We would like you to listen to this audio clip carefully and then answer the questions that follow. \textcolor{gray}{[Present audio clip]}\\

\noindent \textcolor{gray}{[Repeat for all audio clips presented]} Below we are providing a list of broad scenarios of having a conversation with Google Assistant. Select all scenarios that best describe the contents of this audio clip.
$\bigcirc$ Meaningless conversation with Google Assistant
$\bigcirc$ Talking in a funny accent with Google Assistant
$\bigcirc$ Using inappropriate language while talking Google Assistant
$\bigcirc$ Talking with someone in the background while using Google Assistant
$\bigcirc$ Other person using your Google Assistant
$\bigcirc$ Using Google Assistant with other people talking in the background
$\bigcirc$ Using Google Assistant in places with clearly audible background sounds
$\bigcirc$ Accidental activation of Google Assistant
$\bigcirc$ Other\\

\noindent \textcolor{gray}{[Repeat for all audio clips presented]} If you chose ``Other'' in the previous question, please describe the scenario which best describes the contents of this audio clip. Otherwise, please leave this field empty.\_\_\_ \hfill \textcolor{gray}{[Optional]}\\

\noindent \textcolor{gray}{[Repeat for all audio clips presented]} Please choose whether this audio clip was recorded by Google Assistant on a mobile device or voice-enabled Google smart speakers if you recall or know about it?
$\bigcirc$ Google Assistant on a mobile device
$\bigcirc$ Voice-enabled Google smart speakers
$\bigcirc$ I do not recall
$\bigcirc$ I do not know about it
$\bigcirc$ Prefer not to say\\

\noindent \textcolor{gray}{[Repeat for all audio clips presented]} After going through the audio clip, how comfortable would you feel if someone in your intimate relations (such as close friends, lovers, children, and close family members) heard it?
$\bigcirc$ Very comfortable
$\bigcirc$ Slightly comfortable
$\bigcirc$ Neither comfortable nor uncomfortable
$\bigcirc$ Slightly uncomfortable
$\bigcirc$ Very uncomfortable\\

\noindent \textcolor{gray}{[Repeat for all audio clips presented]} After going through the audio clip, how comfortable would you feel if someone in your personal relations (such as friends and associates) heard it?
$\bigcirc$ Very comfortable
$\bigcirc$ Slightly comfortable
$\bigcirc$ Neither comfortable nor uncomfortable
$\bigcirc$ Slightly uncomfortable
$\bigcirc$ Very uncomfortable\\

\noindent \textcolor{gray}{[Repeat for all audio clips presented]} After going through the audio clip, how comfortable would you feel if someone in your social relations (such as strangers and new acquaintances) heard it?
$\bigcirc$ Very comfortable
$\bigcirc$ Slightly comfortable
$\bigcirc$ Neither comfortable nor uncomfortable
$\bigcirc$ Slightly uncomfortable
$\bigcirc$ Very uncomfortable\\

\noindent \textcolor{gray}{[Repeat for all audio clips presented]} After going through the audio clip, how comfortable would you feel if someone in your public relations (such as people you see in public places but don't know) heard it?
$\bigcirc$ Very comfortable
$\bigcirc$ Slightly comfortable
$\bigcirc$ Neither comfortable nor uncomfortable
$\bigcirc$ Slightly uncomfortable
$\bigcirc$ Very uncomfortable\\

\noindent \textcolor{gray}{[Repeat for all audio clips presented]} If Google asked for your consent before collecting this audio clip, how comfortable would you feel sharing it with Google?
$\bigcirc$ Very comfortable
$\bigcirc$ Slightly comfortable
$\bigcirc$ Neither comfortable nor uncomfortable
$\bigcirc$ Slightly uncomfortable
$\bigcirc$ Very uncomfortable\\

\noindent \textcolor{gray}{[Repeat for all audio clips presented]} After going through the audio clip, which of the following privacy policies would you prefer to manage the privacy of similar audio clips in the future?
$\bigcirc$ Give consent to Google to collect and store all such clips indefinitely, until you choose to delete them
$\bigcirc$ Give consent to Google to collect and store all such clips, but delete all stored clips older than 3 months
$\bigcirc$ Give consent to Google to collect and store all such clips, but delete all stored clips older than 18 months
$\bigcirc$ Store all such clips locally on your device, perform necessary computations required by Google on device and share just the computation results instead of complete clips with Google
$\bigcirc$ Do not give consent to Google to collect and store all such clips, and delete any such clips currently stored by Google immediately
$\bigcirc$ None of the above\\

\noindent \textcolor{gray}{[Repeat for all audio clips presented]} Can you please briefly (1-3 sentences) explain your answer choice for the previous question?\_\_\_\\

\noindent \textcolor{gray}{[Repeat for all audio clips presented]} Would you feel comfortable in consenting Google to collect similar audio clips in the future?
$\bigcirc$ Yes
$\bigcirc$ No
$\bigcirc$ I am not sure\\

\noindent \textcolor{gray}{[Repeat for all audio clips presented]} Did you find anything surprising/interesting/shocking about the contents of this audio clip? Please explain your answer briefly (1-3 sentences). Otherwise, you can choose to leave this field empty.\_\_\_ \hfill \textcolor{gray}{[Optional]}\\

\noindent \textcolor{gray}{[Repeat for all commands presented]} The following Google Assistant command was found in your Google My Activity data. We would like you to read this command carefully and then answer the questions that follow. \textcolor{gray}{[Present command]}\\

\noindent \textcolor{gray}{[Repeat for all commands presented]} Below we are providing a list of scenarios of having a conversation with Google Assistant. Select all scenarios that best describe the contents of this Google Assistant command.
$\bigcirc$ Meaningless conversation with Google Assistant
$\bigcirc$ Using inappropriate language while talking Google Assistant
$\bigcirc$ Other person using your Google Assistant
$\bigcirc$ Accidental activation of Google Assistant
$\bigcirc$ Other\\

\noindent \textcolor{gray}{[Repeat for all commands presented]} If you chose ``Other'' in the previous question, please describe the scenario which best describes the contents of this Google Assistant command. Otherwise, please leave this field empty.\_\_\_ \hfill \textcolor{gray}{[Optional]}\\

\noindent \textcolor{gray}{[Repeat for all commands presented]} Please choose whether this command was stored by Google Assistant on a mobile device or voice-enabled Google smart speakers if you recall or know about it?
$\bigcirc$ Google Assistant on a mobile device
$\bigcirc$ Voice-enabled Google smart speakers
$\bigcirc$ I do not recall
$\bigcirc$ I do not know about it
$\bigcirc$ Prefer not to say\\

\noindent \textcolor{gray}{[Repeat for all commands presented]} After going through the Google Assistant command, how comfortable would you feel if someone in your intimate relations (such as close friends, lovers, children, and close family members) came to know about it?
$\bigcirc$ Very comfortable
$\bigcirc$ Slightly comfortable
$\bigcirc$ Neither comfortable nor uncomfortable
$\bigcirc$ Slightly uncomfortable
$\bigcirc$ Very uncomfortable\\

\noindent \textcolor{gray}{[Repeat for all commands presented]} After going through the Google Assistant command, how comfortable would you feel if someone in your personal relations (such as friends and associates) came to know about it?
$\bigcirc$ Very comfortable
$\bigcirc$ Slightly comfortable
$\bigcirc$ Neither comfortable nor uncomfortable
$\bigcirc$ Slightly uncomfortable
$\bigcirc$ Very uncomfortable\\

\noindent \textcolor{gray}{[Repeat for all commands presented]} After going through the Google Assistant command, how comfortable would you feel if someone in your social relations (such as new acquaintances) came to know about it?
$\bigcirc$ Very comfortable
$\bigcirc$ Slightly comfortable
$\bigcirc$ Neither comfortable nor uncomfortable
$\bigcirc$ Slightly uncomfortable
$\bigcirc$ Very uncomfortable\\

\noindent \textcolor{gray}{[Repeat for all commands presented]} After going through the Google Assistant command, how comfortable would you feel if someone in your public relations (such as people you see in public places but don't know) came to know about it?
$\bigcirc$ Very comfortable
$\bigcirc$ Slightly comfortable
$\bigcirc$ Neither comfortable nor uncomfortable
$\bigcirc$ Slightly uncomfortable
$\bigcirc$ Very uncomfortable\\

\noindent \textcolor{gray}{[Repeat for all commands presented]} If Google asked for your consent before storing this Google Assistant command, how comfortable would you feel sharing it with Google?
$\bigcirc$ Very comfortable
$\bigcirc$ Slightly comfortable
$\bigcirc$ Neither comfortable nor uncomfortable
$\bigcirc$ Slightly uncomfortable
$\bigcirc$ Very uncomfortable\\

\noindent \textcolor{gray}{[Repeat for all commands presented]} After going through the Google Assistant command, which of the following privacy policies would you prefer to manage the privacy of similar Google Assistant commands in the future?
$\bigcirc$ Give consent to Google to collect and store all such commands indefinitely, until you choose to delete them
$\bigcirc$ Give consent to Google to collect and store all such commands, but delete all stored commands older than 3 months
$\bigcirc$ Give consent to Google to collect and store all such commands, but delete all stored commands older than 18 months
$\bigcirc$ Store all such commands locally on your device, perform necessary computations required by Google on device and share just the computation results instead of exact commands with Google
$\bigcirc$ Do not give consent to Google to collect and store all such commands, and delete any such commands currently stored by Google immediately
$\bigcirc$ None of the above\\

\noindent \textcolor{gray}{[Repeat for all commands presented]} Can you please briefly (1-3 sentences) explain your answer choice for the previous question?\_\_\_\\

\noindent \textcolor{gray}{[Repeat for all commands presented]} Would you feel comfortable in consenting Google to collect similar commands in the future?
$\bigcirc$ Yes
$\bigcirc$ No
$\bigcirc$ I am not sure\\

\noindent \textcolor{gray}{[Repeat for all commands presented]} Did you find anything surprising/interesting/shocking about the contents of this Google Assistant command? Please explain your answer briefly (1-3 sentences). Otherwise, you can choose to leave this field empty.\_\_\_ \hfill \textcolor{gray}{[Optional]}\\

\noindent The following location was found in your Google My Activity data. We would like you to visit the provided link to learn more about the location and then answer the questions that follow.\\

\noindent Please choose the environment which best describes this location.
$\bigcirc$ Home
$\bigcirc$ Office
$\bigcirc$ Car
$\bigcirc$ Other
$\bigcirc$ I do not recall
$\bigcirc$ I do not know about it
$\bigcirc$ Prefer not to say\\

\noindent If you chose ``Other'' in the previous question, please describe the environment which best describes this location. Otherwise, please leave this field empty.\_\_\_ \hfill \textcolor{gray}{[Optional]}\\

\noindent After going through the location, how comfortable would you feel if someone in your intimate relations (such as close friends, lovers, children, and close family members) came to know about it?
$\bigcirc$ Very comfortable
$\bigcirc$ Slightly comfortable
$\bigcirc$ Neither comfortable nor uncomfortable
$\bigcirc$ Slightly uncomfortable
$\bigcirc$ Very uncomfortable\\

\noindent After going through the location, how comfortable would you feel if someone in your personal relations (such as friends and associates) came to know about it?
$\bigcirc$ Very comfortable
$\bigcirc$ Slightly comfortable
$\bigcirc$ Neither comfortable nor uncomfortable
$\bigcirc$ Slightly uncomfortable
$\bigcirc$ Very uncomfortable\\

\noindent After going through the location, how comfortable would you feel if someone in your social relations (such as new acquaintances) came to know about it?
$\bigcirc$ Very comfortable
$\bigcirc$ Slightly comfortable
$\bigcirc$ Neither comfortable nor uncomfortable
$\bigcirc$ Slightly uncomfortable
$\bigcirc$ Very uncomfortable\\

\noindent After going through the location, how comfortable would you feel if someone in your public relations (such as people you see in public places but don't know) came to know about it?
$\bigcirc$ Very comfortable
$\bigcirc$ Slightly comfortable
$\bigcirc$ Neither comfortable nor uncomfortable
$\bigcirc$ Slightly uncomfortable
$\bigcirc$ Very uncomfortable\\

\noindent If Google asked for your consent before storing this location, how comfortable would you feel sharing it with Google?
$\bigcirc$ Very comfortable
$\bigcirc$ Slightly comfortable
$\bigcirc$ Neither comfortable nor uncomfortable
$\bigcirc$ Slightly uncomfortable
$\bigcirc$ Very uncomfortable\\

\noindent After going through the location, which of the following privacy policies would you prefer to manage the privacy of similar locations in the future?
$\bigcirc$ Give consent to Google to collect and store all such locations indefinitely, until you choose to delete them
$\bigcirc$ Give consent to Google to collect and store all such locations, but delete all stored locations older than 3 months
$\bigcirc$ Give consent to Google to collect and store all such locations, but delete all stored locations older than 18 months
$\bigcirc$ Store all such locations locally on your device, perform necessary computations required by Google on device and share just the computation results instead of approximate locations with Google
$\bigcirc$ Do not give consent to Google to collect and store all such locations, and delete any such locations currently stored by Google immediately
$\bigcirc$ None of the above\\

\noindent Can you please briefly (1-3 sentences) explain your answer choice for the previous question?\_\_\_\\

\noindent Would you feel comfortable in consenting Google to collect similar locations in the future?
$\bigcirc$ Yes
$\bigcirc$ No
$\bigcirc$ I am not sure\\

\noindent Did you find anything surprising/interesting/shocking about this location? Please explain your answer briefly (1-3 sentences). Otherwise, you can choose to leave this field empty.\_\_\_ \hfill \textcolor{gray}{[Optional]}\\

\noindent \textbf{Privacy preferences}

\noindent In general, which of the following privacy policies would you prefer to manage the privacy of your data from Google Assistant on your Android smartphone (e.g., audio clips, commands, etc.) in the future?
$\bigcirc$ Give consent to Google to collect and store all data indefinitely, until you choose to delete it
$\bigcirc$ Give consent to Google to collect and store all data, but delete all data older than 3 months
$\bigcirc$ Give consent to Google to collect and store all data, but delete all data older than 18 months
$\bigcirc$ Store all data locally on your device, perform necessary computations required by Google on device and share just the computation results instead of complete data with Google
$\bigcirc$ Do not give consent to Google to collect and store all data, and delete any data currently stored by Google immediately
$\bigcirc$ None of the above\\

\noindent We would now like you to visit \url{https://myactivity.google.com/myactivity} and check the current status of this option. You can find it by clicking ``Activity controls'' on the left-hand side of the page, and then clicking the ``Auto-delete'' option under the ``Web \& App Activity'' section. Please do not make any changes until the end of this study.\\

\noindent Do you currently have this auto-delete option enabled for your Account?
$\bigcirc$ I have this option enabled (but do not recall enabling it)
$\bigcirc$ I have this option enabled (and recall enabling it)
$\bigcirc$ I do not have this option enabled\\

\noindent Would you prefer to enable this auto-delete option (or keep it enabled) in the future?
$\bigcirc$ Yes
$\bigcirc$ No
$\bigcirc$ I am not sure\\

\noindent On a scale of 1 to 5, with 1 being the least and 5 being the most, how much do you trust Google that it would not use your data for malicious purposes?
$\bigcirc$ 1
$\bigcirc$ 2
$\bigcirc$ 3
$\bigcirc$ 4
$\bigcirc$ 5\\

\noindent For each of the following cases when do you think it would be acceptable for Google to collect and store your data from Google Assistant? \textcolor{gray}{[Matrix-style grid with the following rows]}

\noindent Answer choices for each row-
$\bigcirc$ Always acceptable
$\bigcirc$ Sometimes acceptable
$\bigcirc$ Never acceptable

\begin{itemize}
    \item If you have explicitly consented Google to collect your data
    \item If Google does not notify you while collecting your data
    \item If Google explicitly notifies you while collecting some/all of your data
    \item If Google anonymizes (removes personally identifiable information from) your data before storing it on its data storage facilities
    \item If you get financial rewards (in the form of coupon discounts, gift cards, etc.) for sharing your data with Google
    \item If your data is used to improve the performance of Google Assistant software
    \item If your data is used to develop new features for Google Assistant software
    \item If your data is stored indefinitely by Google on its data storage facilities
    \item If your data is not stored indefinitely by Google on its data storage facilities
\end{itemize}

\noindent At this stage of the study, we would like to inform you that the data we presented to you in earlier sections was automatically flagged by our system as potentially sensitive due to several reasons. We have prepared a brief explanation for each of the data points presented. Please go through the data presented in each category one by one and then read the explanation provided. \textcolor{gray}{[Provide explanation for why each data element was presented]}\\

\noindent \textcolor{gray}{[Repeat for all data elements presented]} Based on how relevant you find a data point to the explanation provided, rate the accuracy of each classification on a scale of 1 to 5, with 1 being the least and 5 being the most.
$\bigcirc$ 1
$\bigcirc$ 2
$\bigcirc$ 3
$\bigcirc$ 4
$\bigcirc$ 5\\

\noindent We would now like you to visit the following Google Drive folder which contains all your data that was flagged by our system as potentially sensitive. For your convenience, we have classified the data into several categories, which further have sub-categories. The division of categories and sub-categories has been done intuitively as explained in the table above. The names of files, folders and sub-folders in the Google Drive folder indicate these categories and sub-categories. There are three explanation files placed in the Google Drive folder that will help you to understand why a particular file was flagged by the system.

\begin{enumerate}[label=(\Alph*)]
    \item EXPLANATION FOR AUDIO(PART-I).txt- This file contains explanations for all data points that have a companion audio file for 10 categories- NOISY BACKGROUND(S), MULTIPLE SPEAKERS/BREAKS, 2 categories of speakers (MALE SPEAKER(S)/FEMALE SPEAKER(S)), and 6 categories of background noises (WIND/RAIN/OCEAN/CHATTER/TRAFFIC/\\HIGHWAY).
    \item EXPLANATION FOR AUDIO(PART-II).txt - This file contains explanations for all data points that have a companion audio file for 4 categories- GRAMMATICAL ERROR(S), NON-STANDARD WORD(S), REGRET WORD(S), and NEGATIVE SENTIMENT.
    \item EXPLANATION FOR COMMANDS.txt- This file contains explanations for all data points that have only commands (but no companion audio file) for 4 categories- GRAMMATICAL ERROR(S), NON-STANDARD WORD(S), REGRET WORD(S), and NEGATIVE SENTIMENT.
\end{enumerate}

\noindent Note that some of these files/folders might not be present in the Google Drive folder if no data was found for the corresponding category in your Google-collected activity data. We would like you to go through some of the data presented (around 10 files) in the Google Drive folder. You might have to download the presented audio files in order to listen to them. After you go through the data, we will ask you a few questions related to it. You may also download the Google Drive folder for your reference. \textcolor{gray}{[Provide link to the personalized Google Drive folder]}\\

\noindent Among the data presented across several categories and sub-categories, which ones did you check out? Select all that apply.
$\square$ AUDIO CLIPS WITH NOISY BACKGROUND(S)
$\square$ AUDIO CLIPS WITH MULTIPLE SPEAKERS/BREAKS
$\square$ AUDIO CLIPS WITH MALE SPEAKER(S)
$\square$ AUDIO CLIPS WITH FEMALE SPEAKER(S)
$\square$ AUDIO CLIPS WITH WIND BACKGROUND(S)
$\square$ AUDIO CLIPS WITH RAIN BACKGROUND(S)
$\square$ AUDIO CLIPS WITH OCEAN BACKGROUND(S)
$\square$ AUDIO CLIPS WITH TRAFFIC BACKGROUND(S)
$\square$ AUDIO CLIPS WITH HIGHWAY BACKGROUND(S)
$\square$ AUDIO CLIPS WITH CHATTER BACKGROUND(S)
$\square$ AUDIO CLIPS WITH GRAMMATICAL ERROR(S)
$\square$ AUDIO CLIPS WITH NON-STANDARD WORD(S)
$\square$ AUDIO CLIPS WITH REGRET WORD(S)
$\square$ AUDIO CLIPS WITH NEGATIVE SENTIMENT
$\square$ TEXTUAL COMMANDS WITH GRAMMATICAL ERROR(S)
$\square$ TEXTUAL COMMANDS WITH NON-STANDARD WORD(S)
$\square$ TEXTUAL COMMANDS WITH REGRET WORD(S)
$\square$ TEXTUAL COMMANDS WITH NEGATIVE SENTIMENT\\

\noindent On a scale of 1 to 5, with 1 being not sensitive and 5 very sensitive, how sensitive did you find the data, on average?
$\bigcirc$ 1
$\bigcirc$ 2
$\bigcirc$ 3
$\bigcirc$ 4
$\bigcirc$ 5\\

\noindent On a scale of 1 to 5, with 1 being very inaccurate and 5 very accurate, how accurate did you find the classification of these potentially sensitive data by our system?
$\bigcirc$ 1
$\bigcirc$ 2
$\bigcirc$ 3
$\bigcirc$ 4
$\bigcirc$ 5\\

\noindent While going through the potentially sensitive data, did you come across at least some data that you did not expect to be present in your Google My Activity data?
$\bigcirc$ Yes
$\bigcirc$ No
$\bigcirc$ I am not sure\\

\noindent If you found some data that especially caught your attention, please briefly (1-3 sentences) explain why it caught your attention. Otherwise, you can choose to leave this field empty.\_\_\_ \hfill \textcolor{gray}{[Optional]}\\

\noindent In general, which of the following privacy policies would you prefer to manage the privacy of your data from Google Assistant on your Android smartphone (e.g., audio clips, commands, etc.) in the future?
$\bigcirc$ Give consent to Google to collect and store all data indefinitely, until you choose to delete it
$\bigcirc$ Give consent to Google to collect and store all data, but delete all data older than 3 months
$\bigcirc$ Give consent to Google to collect and store all data, but delete all data older than 18 months
$\bigcirc$ Store all data locally on your device, perform necessary computations required by Google on device and share just the computation results instead of complete data with Google
$\bigcirc$ Do not give consent to Google to collect and store all data, and delete any data currently stored by Google immediately
$\bigcirc$ None of the above\\

\noindent Google provides users with an auto-delete option that allows them to automatically delete activity older than 3 months/18 months. Would you prefer to enable this auto-delete option (or keep it enabled) in the future?
$\bigcirc$ Yes
$\bigcirc$ No
$\bigcirc$ I am not sure\\

\noindent On a scale of 1 to 5, with 1 being the least and 5 being the most, how much do you trust Google that it would not use your data for malicious purposes, after participating in this study?
$\bigcirc$ 1
$\bigcirc$ 2
$\bigcirc$ 3
$\bigcirc$ 4
$\bigcirc$ 5\\

\noindent For each of the following cases when do you think it would be acceptable for Google to collect and store your data from Google Assistant? We asked you these questions in an earlier section. We want to see if you would like to change some of your responses after checking some of the files in the Google Drive folder. \textcolor{gray}{[Matrix-style grid with the following rows]}

\noindent Answer choices for each row-
$\bigcirc$ Always acceptable
$\bigcirc$ Sometimes acceptable
$\bigcirc$ Never acceptable

\begin{itemize}
    \item If you have explicitly consented Google to collect your data
    \item If Google does not notify you while collecting your data
    \item If Google explicitly notifies you while collecting some/all of your data
    \item If Google anonymizes (removes personally identifiable information from) your data before storing it on its data storage facilities
    \item If you get financial rewards (in the form of coupon discounts, gift cards, etc.) for sharing your data with Google
    \item If your data is used to improve the performance of Google Assistant software
    \item If your data is used to develop new features for Google Assistant software
    \item If your data is stored indefinitely by Google on its data storage facilities
    \item If your data is not stored indefinitely by Google on its data storage facilities
\end{itemize}

\noindent \textbf{Awareness}

\noindent Compared to before participating in this study, how likely are you to read about the privacy policies of collected data of other voice assistants (you use or will use in the future)?
$\bigcirc$ Very likely
$\bigcirc$ Somewhat likely
$\bigcirc$ Not likely\\

\noindent Compared to before participating in this study, how likely are you to learn more about the data collection practices of other voice assistants before using them?
$\bigcirc$ Very likely
$\bigcirc$ Somewhat likely
$\bigcirc$ Not likely\\

\noindent Compared to before participating in this study, how likely are you to delete some of your data collected by Google?
$\bigcirc$ Very likely
$\bigcirc$ Somewhat likely
$\bigcirc$ Not likely\\

\noindent Would you like to visit your Google My Activity dashboard more often in the future?
$\bigcirc$ Yes
$\bigcirc$ No
$\bigcirc$ I am not sure\\

\noindent Please briefly (1-3 sentences) explain what new things you learnt about Google Assistant by participating in this study?\_\_\_\\

\noindent Do you feel that tech companies, in general, tend to make it harder for users to access their collected data?
$\bigcirc$ Yes
$\bigcirc$ No
$\bigcirc$ I am not sure\\

\noindent \textbf{General privacy perception related questions}

\noindent On a scale of 1 to 5, with 1 being strongly disagree and 5 being strongly agree, express your level of agreement for each of the statements given below. Here 1 indicates ``Strongly disagree'', and 5 indicates ``Strongly agree''.
\textcolor{gray}{[Matrix-style grid with the following rows]}

\noindent Answer choices for each row-
$\bigcirc$ 1
$\bigcirc$ 2
$\bigcirc$ 3
$\bigcirc$ 4
$\bigcirc$ 5

\begin{itemize}
    \item Companies seeking information online should disclose the way the data are collected, processed, and used.
    \item A good consumer online privacy policy should have a clear and conspicuous disclosure.
    \item It is very important to me that I am aware and knowledgeable about how my personal information will be used.
    \item It usually bothers me when online companies ask me for personal information.
    \item When online companies ask me for personal information, I sometimes think twice before providing it.
    \item It bothers me to give personal information to so many online companies.
    \item I'm concerned that online companies are collecting too much personal information about me.
    \item I feel that I should have more control over how online companies collect and use my personal information.
\end{itemize}

\noindent \textbf{Concluding thoughts}

\noindent Do you think that, in general, our sensitive content detection system provided valuable assistance in finding sensitive data on Google servers?
$\bigcirc$ Yes
$\bigcirc$ No
$\bigcirc$ I am not sure\\

\noindent If we made this sensitive content detection system publicly available, would you recommend others to try it out?
$\bigcirc$ Yes
$\bigcirc$ No
$\bigcirc$ I am not sure\\

\noindent Google uses audio clips collected from users to develop and improve its audio recognition technologies and the Google services that use them, like Google Assistant. If Google used a similar sensitive content detection system to filter out all sensitive audio clips before storing your audio clip data, would you consider giving consent to Google to collect and store some of your non-sensitive audio clip data?
$\bigcirc$ Yes
$\bigcirc$ No
$\bigcirc$ I am not sure\\

\noindent Do you think that this study made you more aware of the data collection practices of Google Assistant?
$\bigcirc$ Yes
$\bigcirc$ No
$\bigcirc$ I am not sure\\

\noindent Did you enjoy participating in this study?
$\bigcirc$ Yes
$\bigcirc$ No
$\bigcirc$ I am not sure\\

\noindent Do you have any suggestions for the final design of this sensitive content detection system? Please enter them in the text box provided below.\_\_\_ \hfill \textcolor{gray}{[Optional]}

\end{scriptsize}

\section{\newtext{Other Settings of Recommendation System}}\label{sec:maybevariations}
\begin{figure}[!t]
\centering
\includegraphics[width=\linewidth]{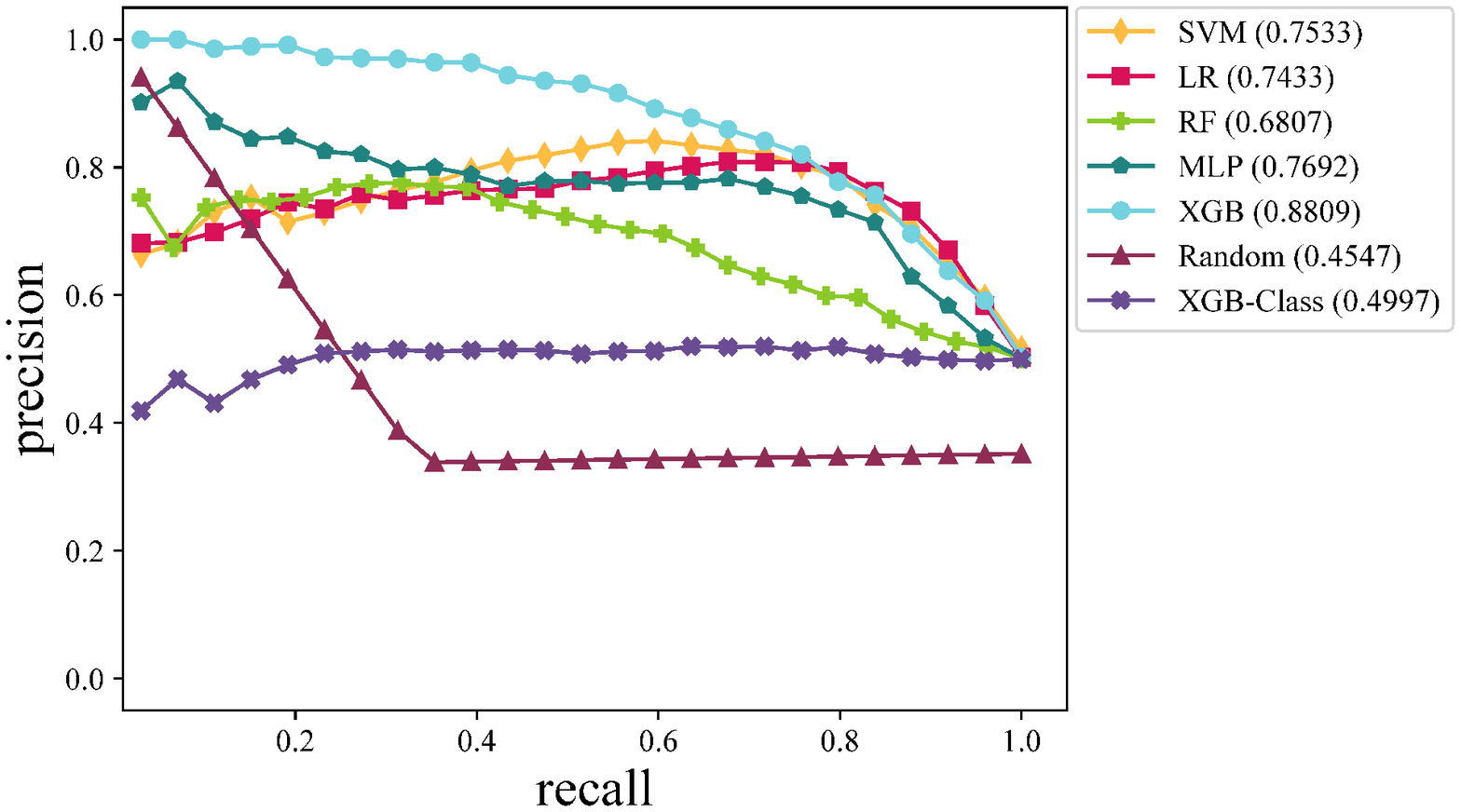}
\caption{\newtext{PR curves while classifying data elements for `Yes' setting (SVM, LR, RF, MLP, and XGB are evaluated ML models, whereas Random and XGB-Class are baseline models)}}
\label{fig:dm_sep_yes}
\end{figure}

\begin{figure}[!t]
\centering
\includegraphics[width=\linewidth]{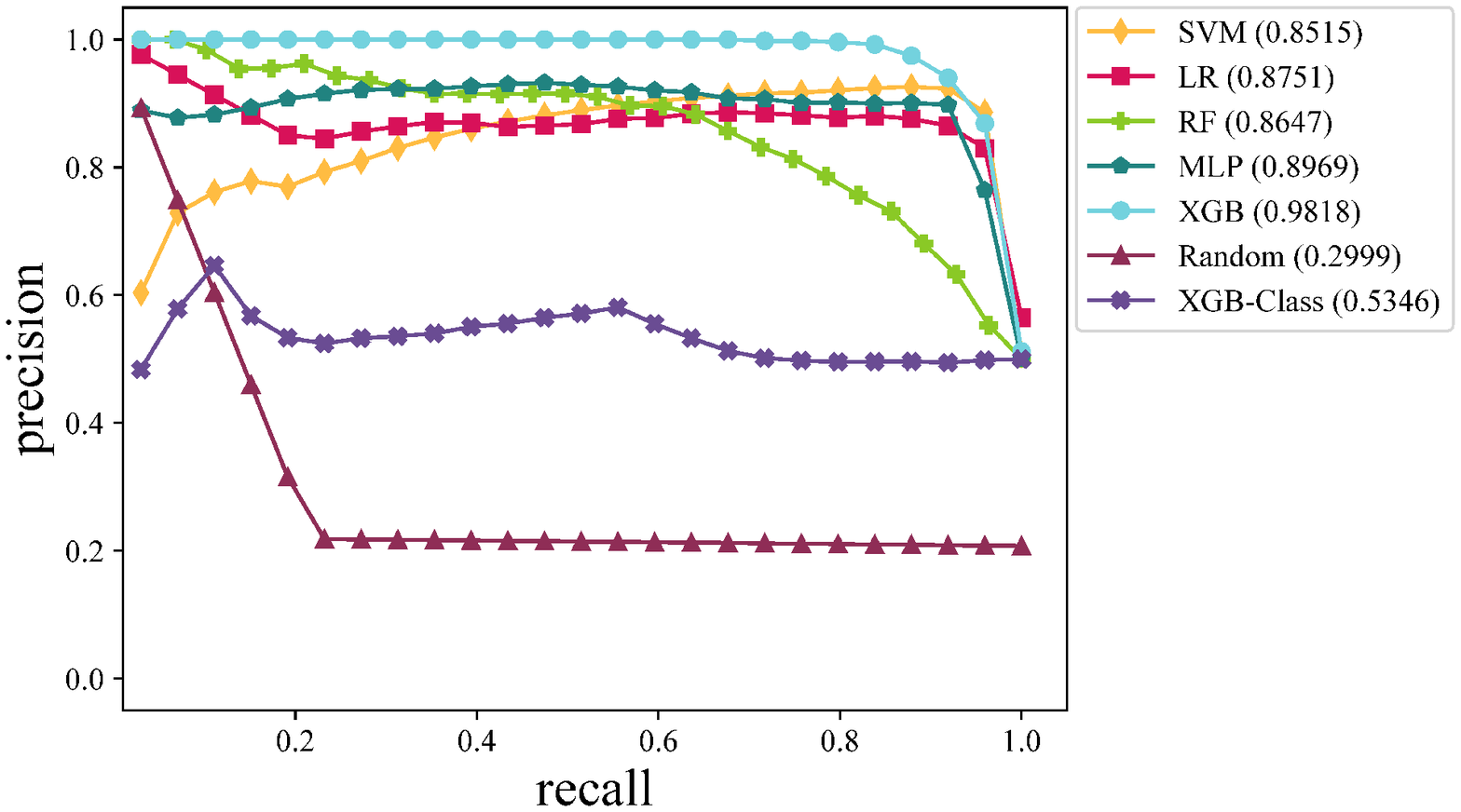}
\caption{\newtext{PR curves while classifying data elements for `Removed' setting (SVM, LR, RF, MLP, and XGB are evaluated ML models, whereas Random and XGB-Class are baseline models)}}
\label{fig:dm_sep_leave}
\end{figure}

\begin{figure}[!t]
\centering
\includegraphics[width=\linewidth]{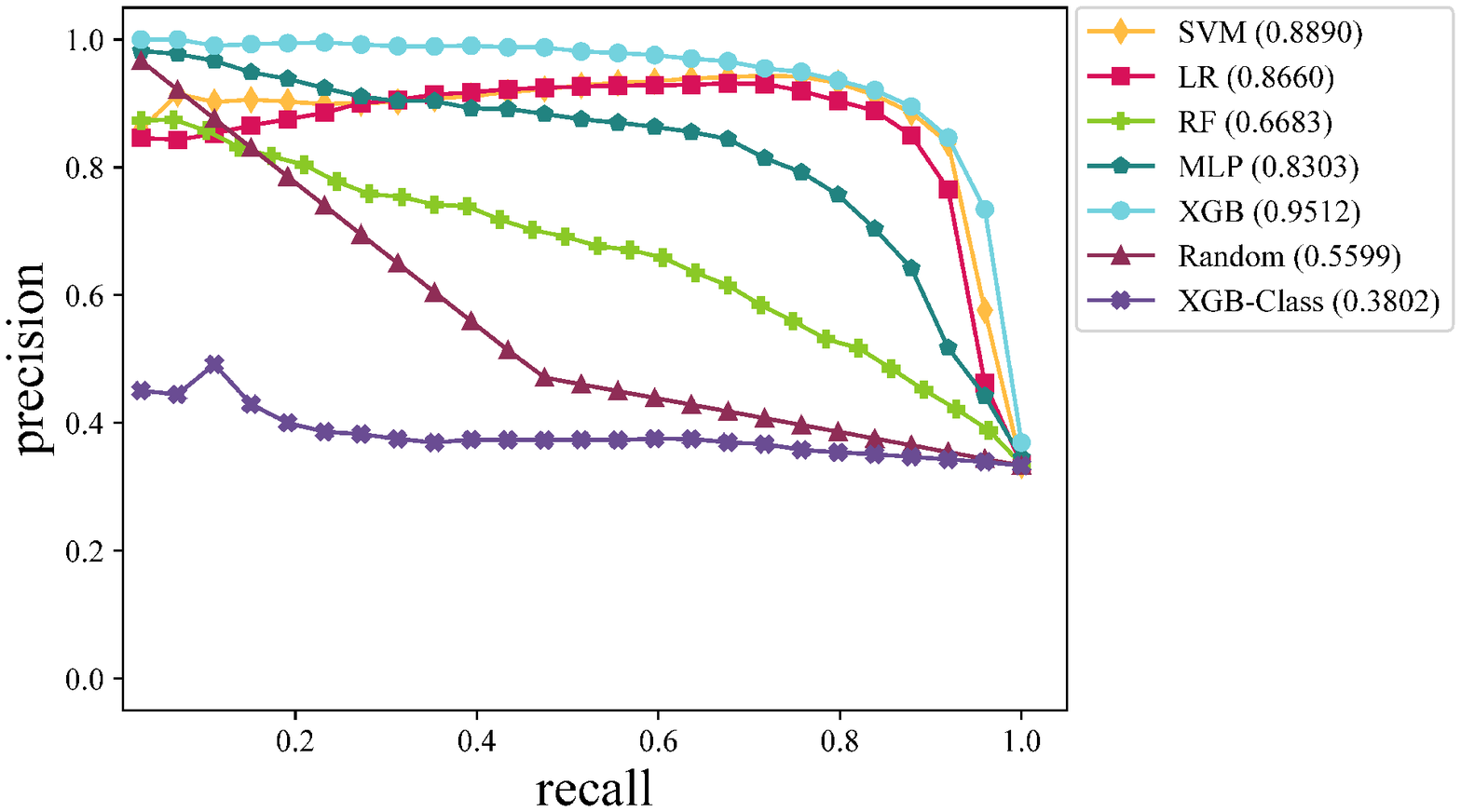}
\caption{\newtext{PR curves while classifying data elements for `Separate' setting (SVM, LR, RF, MLP, and XGB are evaluated ML models, whereas Random and XGB-Class are baseline models)}}
\label{fig:dm_sep_separate}
\end{figure}

\begin{figure}[!t]
\centering
\includegraphics[width=0.9\linewidth]{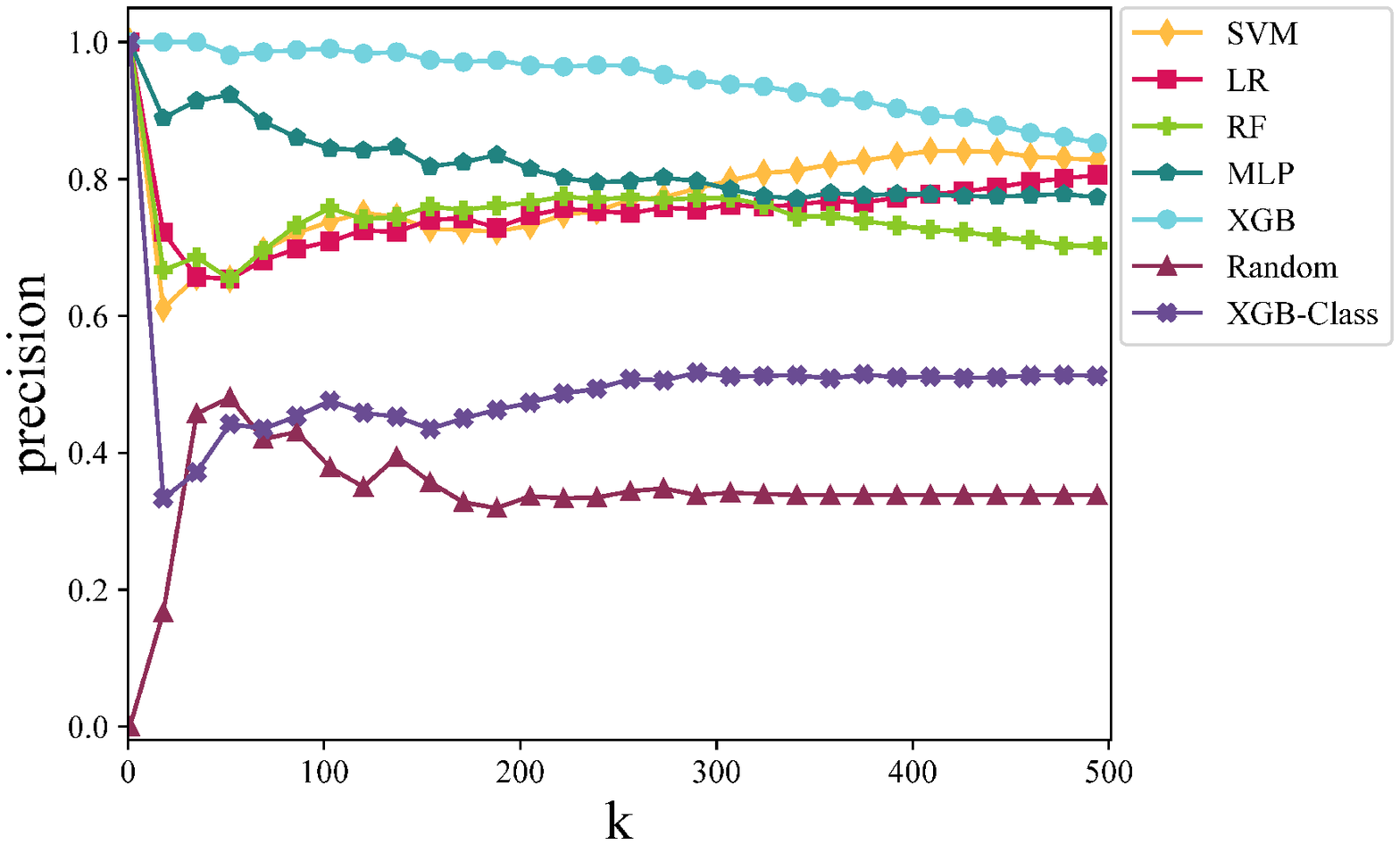}
\caption{\newtext{Precision@k curves while classifying data elements for `Yes' setting (SVM, LR, RF, MLP, and XGB are evaluated ML models, whereas Random and XGB-Class are baseline models)}}
\label{fig:P@K_yes}
\end{figure}

\begin{figure}[!t]
\centering
\includegraphics[width=0.9\linewidth]{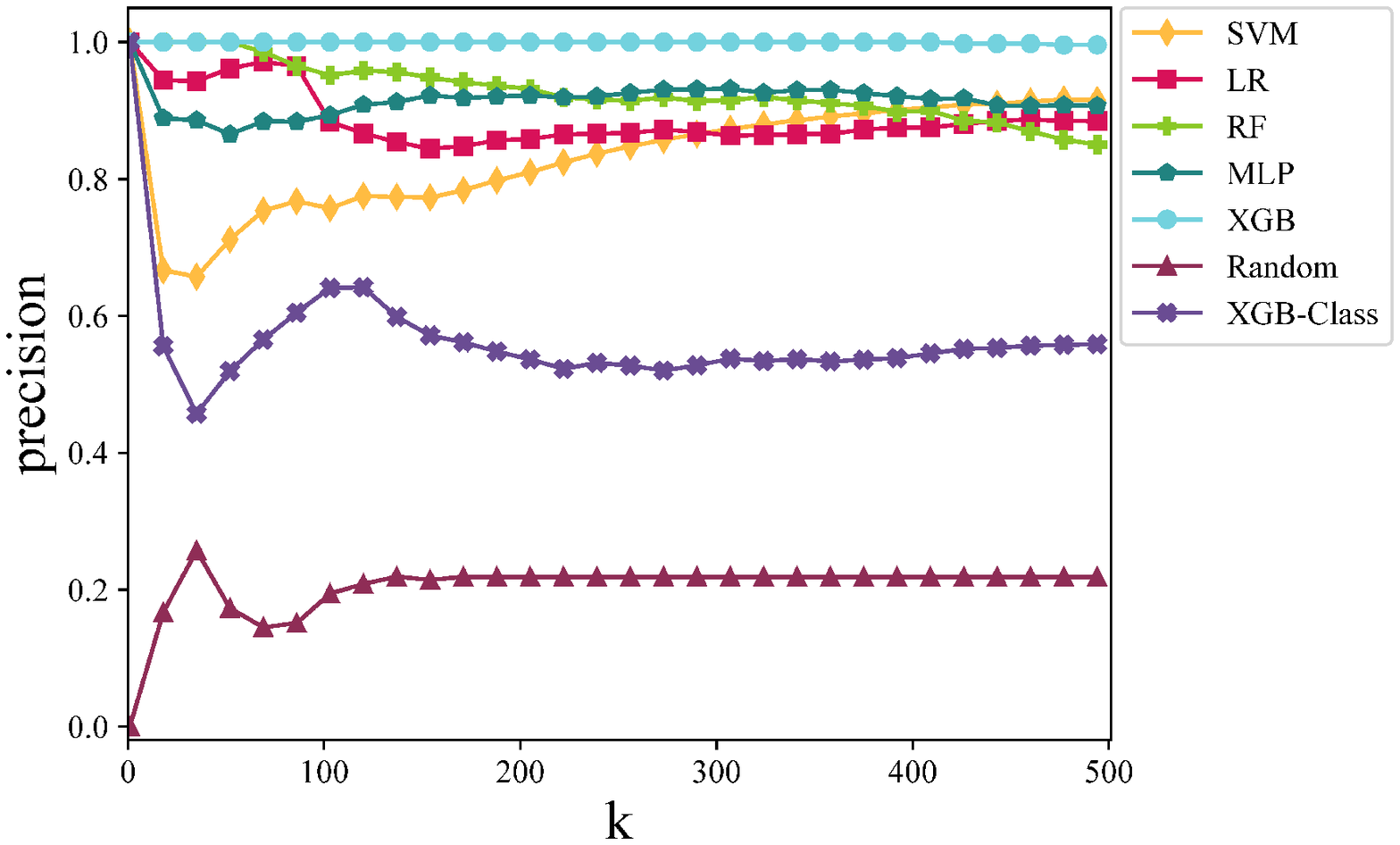}
\caption{\newtext{Precision@k curves while classifying data elements for `Removed' setting (SVM, LR, RF, MLP, and XGB are evaluated ML models, whereas Random and XGB-Class are baseline models)}}
\label{fig:P@K_leave}
\end{figure}

\begin{figure}[!t]
\centering
\includegraphics[width=0.9\linewidth]{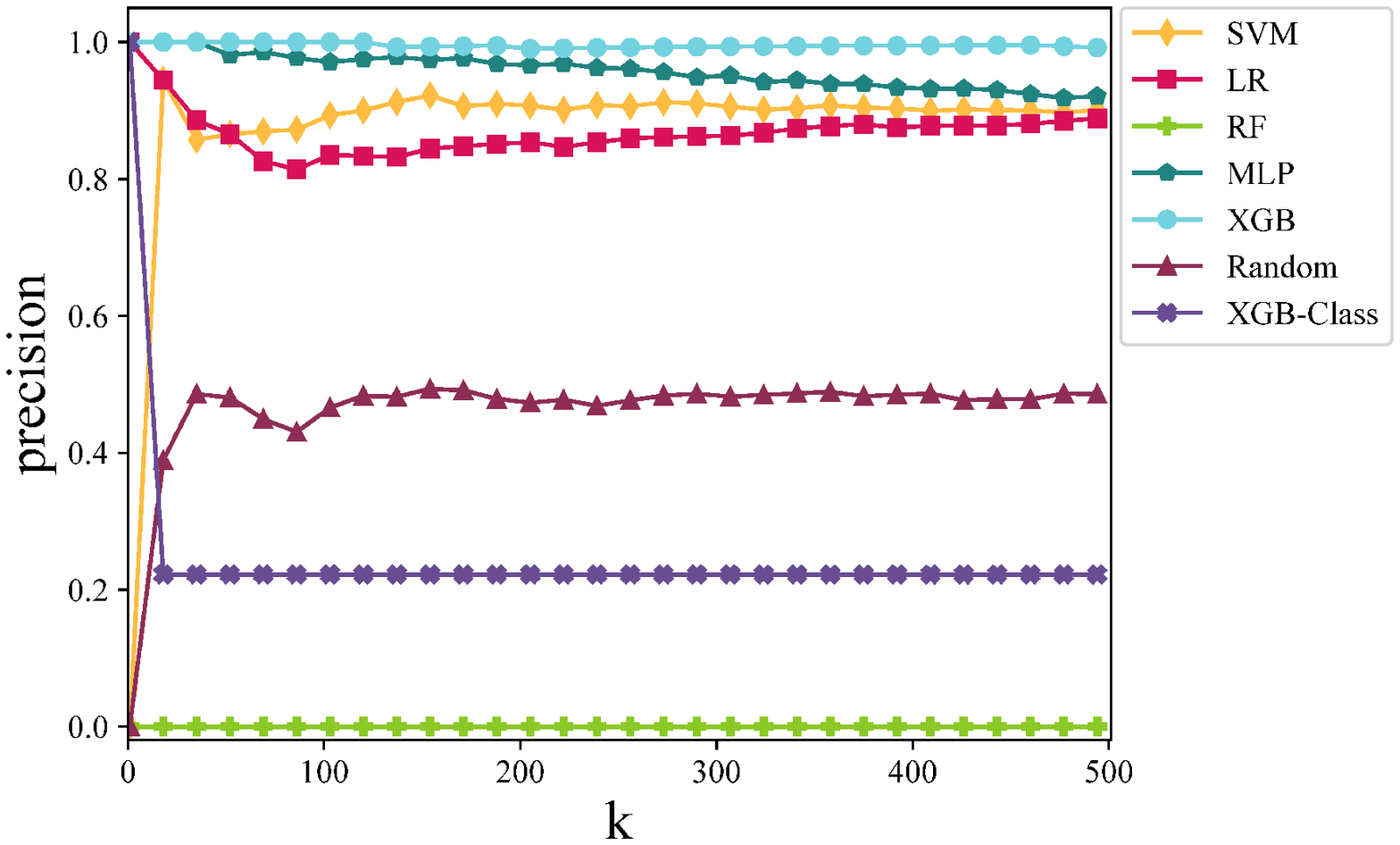}
\caption{\newtext{Precision@k curves while classifying data elements for `Separate' setting (SVM, LR, RF, MLP, and XGB are evaluated ML models, whereas Random and XGB-Class are baseline models)}}
\label{fig:P@K_separate}
\end{figure}

\newtext{Tables~\ref{tab:result_yes},~\ref{tab:result_leave}, and~\ref{tab:result_separate} show the macro-averaged precision, recall, and F1 scores for all models for other settings of the ML-based recommendation system---(a) `Yes'--When the `I am not sure' label is treated as a proxy for the `Yes' label, (b) `Removed'--When the `I am not sure' label is not considered in training process, and (c) `Separate'--When the `I am not sure' label is treated as a separate class, respectively. Figures~\ref{fig:dm_sep_yes},~\ref{fig:dm_sep_leave}, and~\ref{fig:dm_sep_separate} show the PR curves generated for all models for the above-mentioned settings, respectively. Similarly, Figures~\ref{fig:P@K_yes},~\ref{fig:P@K_leave}, and~\ref{fig:P@K_separate} show the precision@k curves generated for all models for the three settings, respectively.}

\begin{table}[!t]
\small
    \newtext{
    \small
    \centering
    \begin{tabularx}{\linewidth}{|Y|Y|Y|Y|}
    \hline
    \bf Model & \bf Precision & \bf Recall & \bf F1 score \\ \hline
    \multicolumn{4}{|c|}{\bf Proposed Feature-based Models} \\\hline
    SVM & 0.78 & 0.77 & 0.77 \\ \hline
    LR & \textbf{0.80} & \textbf{0.79} & \textbf{0.79}\\ \hline
    RF & 0.67  & 0.67 & 0.67\\ \hline
    MLP & 0.76  & 0.76 & 0.76 \\ \hline
    XGB & 0.79 & \textbf{0.79}  & \textbf{0.79} \\ \hline
    \multicolumn{4}{|c|}{\bf Baseline Models} \\ \hline
    Random & 0.49  & 0.49 & 0.49\\ \hline
    XGB-Class & 0.52 & 0.52 & 0.51\\ \hline
    \end{tabularx}
    \caption{Macro-averaged Precision, Recall, F1-score for all models for `Yes' setting. The highest values in each column are boldfaced.}
    \label{tab:result_yes}
    }
\end{table}

\begin{table}[!t]
\small
    \newtext{
    \small
    \centering
    \begin{tabularx}{\linewidth}{|Y|Y|Y|Y|}
    \hline
    \bf Model & \bf Precision & \bf Recall & \bf F1 score \\ \hline
    \multicolumn{4}{|c|}{\bf Proposed Feature-based Models} \\\hline
    SVM & 0.91 & 0.90 & 0.90 \\ \hline
    LR & 0.90  & 0.90 & 0.90\\ \hline
    RF & 0.78  & 0.78 & 0.78\\ \hline
    MLP & 0.90  & 0.90 & 0.90 \\ \hline
    XGB & \textbf{0.93} & \textbf{0.93}  & \textbf{0.93} \\ \hline
    \multicolumn{4}{|c|}{\bf Baseline Models} \\ \hline
    Random & 0.51  & 0.51 & 0.51\\ \hline
    XGB-Class & 0.58 & 0.58 & 0.58\\ \hline
    \end{tabularx}
    \caption{Macro-averaged Precision, Recall, F1-score for all models for `Removed' setting. The highest values in each column are boldfaced.}
    \label{tab:result_leave}
    }
\end{table}

\begin{table}[!t]
\small
    \newtext{
    \small
    \centering
    \begin{tabularx}{\linewidth}{|Y|Y|Y|Y|}
    \hline
    \bf Model & \bf Precision & \bf Recall & \bf F1 score \\ \hline
    \multicolumn{4}{|c|}{\bf Proposed Feature-based Models} \\\hline
    SVM & 0.87 & 0.86 & 0.85 \\ \hline
    LR & 0.87  & 0.86 & 0.85\\ \hline
    RF & 0.64  & 0.63 & 0.63\\ \hline
    MLP & 0.75  & 0.75 & 0.74 \\ \hline
    XGB & \textbf{0.89} & \textbf{0.89}  & \textbf{0.89} \\ \hline
    \multicolumn{4}{|c|}{\bf Baseline Models} \\ \hline
    Random & 0.32  & 0.32 & 0.32\\ \hline
    XGB-Class & 0.38 & 0.38 & 0.38\\ \hline
    \end{tabularx}
    \caption{Macro-averaged Precision, Recall, F1-score for all models for `Separate' setting. The highest values in each column are boldfaced.}
    \label{tab:result_separate}
    }
\end{table}

\section{Additional Tables and Figures}
\label{sec:additional}

This appendix presents some additional tables and figures.

\begin{table}[!t]
    \footnotesize
    \centering
    \begin{tabularx}{\linewidth}{|c|Y|Y|Y|Y|}
        \hline
        \multirow{2}{*}{\textbf{Class}} & \multicolumn{4}{c|}{\textbf{Proxemic zone}}\\
        \cline{2-5}
         & \textbf{Intimate} & \textbf{Private} & \textbf{Social} & \textbf{Public}\\
        \hline
        audio-noise & \textbf{0.780} & 0.925 & 0.949 & 0.846\\ \hline
        audio-multi-spkr & 0.891 & 0.889 & 0.886 & \textbf{0.821}\\ \hline
        audio-non-gend & 0.960 & 0.961 & 0.961 & \textbf{0.859}\\ \hline
        audio-non-bkgd & 0.958 & 0.958 & 0.979 & \textbf{0.852}\\ \hline
        audio-grammar & 0.953 & 0.976 & \textbf{0.724} & 0.774\\ \hline
        audio-non-eng & 0.982 & \textbf{0.966} & 0.982 & 0.982\\ \hline
        audio-regret & 0.855 & 0.877 & 0.800 & \textbf{0.795}\\ \hline
        audio-neg-sent & 0.962 & 0.962 & 0.855 & \textbf{0.803}\\ \hline
        audio-rand & \textbf{0.870} & 0.903 & 0.903 & 0.903\\ \hline
        transcript-typo & 0.929 & 0.929 & 0.950 & \textbf{0.749}\\ \hline
        transcript-non-eng & 0.915 & 0.936 & 0.821 & \textbf{0.762}\\ \hline
        transcript-regret & 0.932 & 0.953 & 0.898 & \textbf{0.832}\\ \hline
        transcript-neg-sent & 0.819 & 0.819 & 0.712 & \textbf{0.710}\\ \hline
        transcript-rand & 0.900 & 0.915 & \textbf{0.829} & 0.872\\ \hline
        location & 0.771 & 0.845 & 0.799 & \textbf{0.626}\\
        \hline
    \end{tabularx}
    \caption{ Effect size for association between different proxemic zones and Google for each class (Table~\ref{tab:classes}). The lowest values in each row are boldfaced. All associations were statistically significant (Fisher's exact, \textit{p}~$<$~0.05).}
    \label{tab:effect}
\end{table}

\begin{table*}[!t]
    \small
    \centering
    \newtext{
    \begin{tabularx}{\linewidth}{|Y|Y|S[table-format=1.0]|S[table-format=3.0]|S[table-format=4.0]|S[table-format=4.0]|S[table-format=5.1]|S[table-format=5.1]|S[table-format=5.1]|S[table-format=5.0]|}
        \hline
        \multicolumn{2}{|c|}{\textbf{Data element}} & \textbf{\phantom{x}2013\phantom{x}} & \textbf{\phantom{x}2014\phantom{x}} & \textbf{\phantom{x}2015\phantom{x}} & \textbf{\phantom{x}2016\phantom{x}} & \textbf{\phantom{.1}2017} & \textbf{\phantom{.1}2018} & \textbf{\phantom{.1}2019} & \textbf{\phantom{1}2020\phantom{1}}\\
        \hline
        \multirow{4}{*}{\# Audio w/ transcript} & \textbf{Total} & 0 & 902 & 3736 & 5910 & 13809 & 17374 & 22015 & 19889 \\ \cline{2-10}
        & \textbf{Min.} & 0 & 0 & 0 & 0 & 0 & 0 & 0 & 0\\ \cline{2-10}
        & \textbf{Median} & 0 & 0 & 0 & 0 & 2.5 & 31.5 & 77.5 & 46\\ \cline{2-10}
        & \textbf{Max.} & 0 & 296 & 633 & 846 & 3250 & 3585 & 5586 & 6415\\ \hline
        \multirow{4}{*}{\# Only transcripts} & \textbf{Total} & 2 & 0 & 0 & 1668 & 7005 & 8383 & 16774 & 21411 \\ \cline{2-10}
        & \textbf{Min.} & 0 & 0 & 0 & 0 & 0 & 0 & 0 & 0 \\ \cline{2-10}
        & \textbf{Median} & 0 & 0 & 0 & 0 & 0 & 29.5 & 57 & 28\\ \cline{2-10}
        & \textbf{Max.} & 2 & 0 & 0 & 613 & 816 & 775 & 1836 & 3188\\ \hline
        \multirow{4}{*}{\# Ambient location} & \textbf{Total} & 1 & 0 & 0 & 1517 & 5787 & 15061 & 28980 & 32963\\ \cline{2-10}
        & \textbf{Min.} & 0 & 0 & 0 & 0 & 0 & 0 & 0 & 0\\ \cline{2-10}
        & \textbf{Median} & 0 & 0 & 0 & 0 & 0 & 22.5 & 104.5 & 62 \\ \cline{2-10}
        & \textbf{Max.} & 1 & 0 & 0 & 472 & 722 & 2875 & 4124 & 4606 \\ \hline
        \multicolumn{2}{|c|}{Total \# data elements} & 2 & 902 & 3736 & 7578 & 20814 & 25757 & 38789 & 41300\\ \hline
    \end{tabularx}
    \caption{Year-wise overview of participants' GVA data.}
    }
    \label{tab:yearwise_data}
\end{table*}

\begin{figure*}[!t]
\centering
\includegraphics[width=\linewidth]{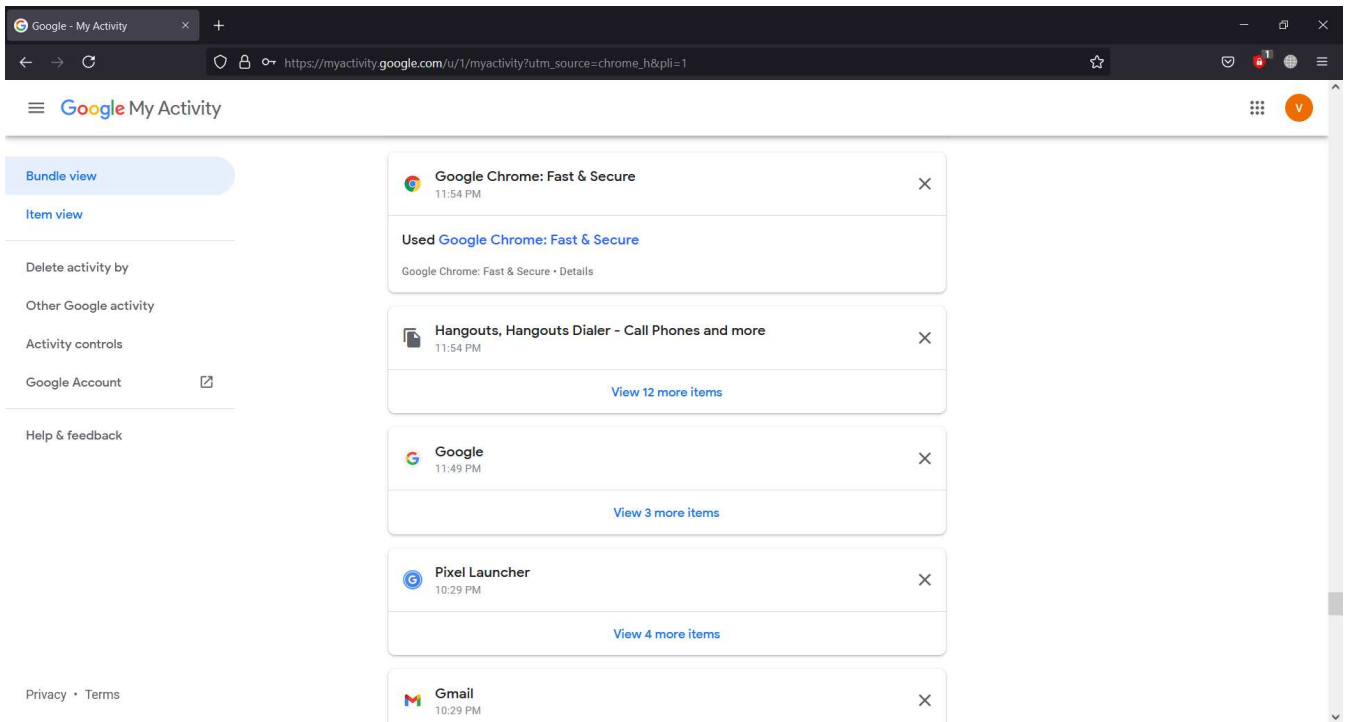}
\caption{User interface of the Google My Activity Dashboard (Last Accessed on June 2021)}
\label{fig:myactivity}
\end{figure*}

\begin{figure*}[!t]
\centering
\includegraphics[width=\linewidth]{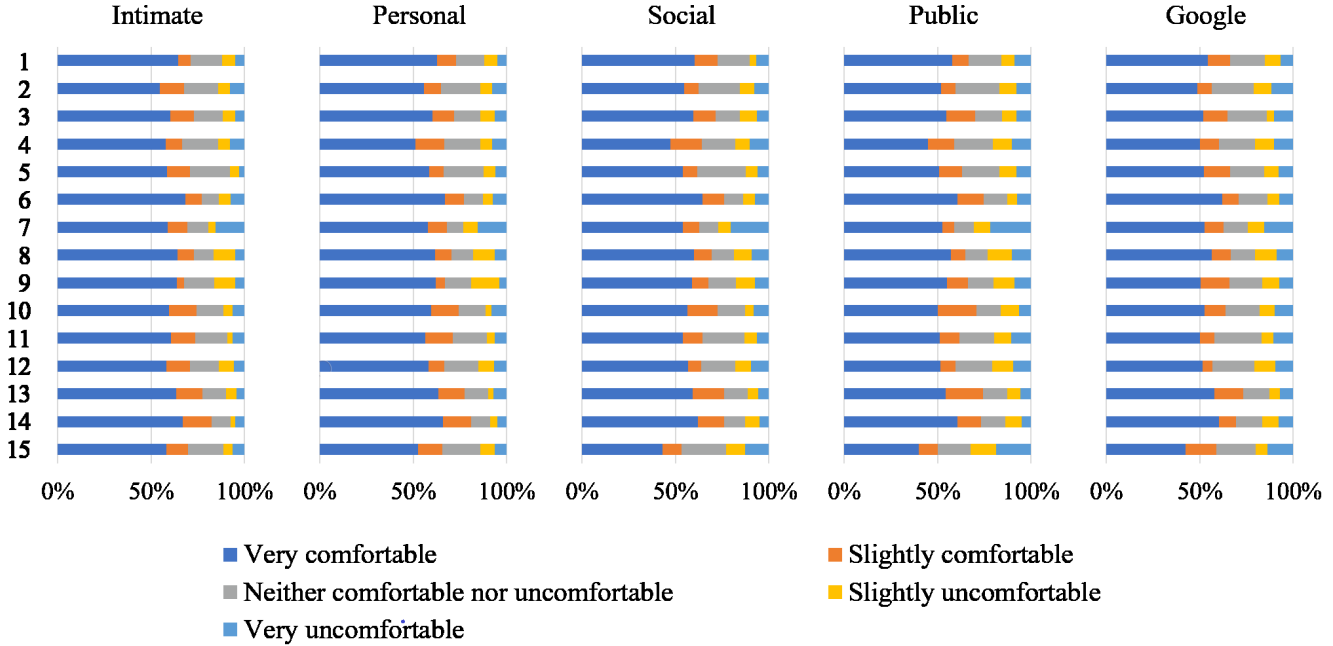}
\caption{User preferences for sharing data elements in each class (Table~\ref{tab:classes}) with people in four proxemic zones and Google}
\label{fig:comfortability}
\end{figure*}

\begin{figure}[!t]
\centering
\includegraphics[width=0.9\linewidth]{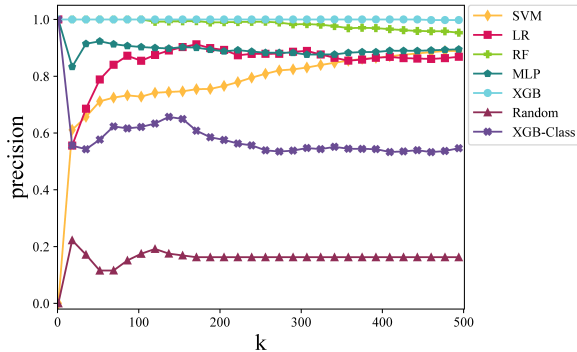}
\caption{Precision@k curve while classifying data elements (SVM, LR, RF, MLP, and XGB are proposed feature-based ML models, whereas Random and XGB-Class are baseline models)}
\label{fig:P@K_500}
\end{figure}

\section{Basic Data Classification}
\label{subsec:basic}

\begin{table}[!t]
    \centering
	\footnotesize
    \begin{tabularx}{\linewidth}{|c|X|}
        \hline
        \textbf{Serial} & \textbf{Description of Scenarios}\\
        \hline
        1 & Talking in a funny accent with GVA\\ \hline
        2 & Small talk with GVA\\ \hline
        3 & Using inappropriate language while talking to GVA\\ \hline
        4 & Talking with someone while also using GVA\\\hline
        5 & Other person using your GVA\\\hline
        6 & Using GVA with other people talking in the background\\\hline
        7 & Using GVA in places with audible background sounds\\\hline
        8 & Accidental activation of GVA\\
        \hline
    \end{tabularx}
    \caption{List of possible usage GVA usage scenarios we considered for uncovering experiences with possibly sensitive data collected by GVA.}
    \label{tab:scenarios}
\end{table}

We identified (and leveraged in Survey~1)  a set of scenarios where according to earlier work, potentially sensitive data might be collected by GVA~\cite{malkin2019attitudes, bentley2018usage, yuting2019ipa}. These scenarios are presented in Table~\ref{tab:scenarios}. Based on these scenarios we created 15 classes and aimed to label data elements with these classes. Specifically, we created twelve small binary classifiers (audio-rand, transcript-rand, and location classes does not require any classifier), four of them classifying audio clips using Signal Processing techniques, and eight of them classifying transcripts (four each for transcripts w/ and w/o companion audio clips) using Natural Language Processing (NLP) techniques. We now explain these classifiers in detail.

\subsection{Signal Processing} We used MATLAB and the MATLAB Audio Analysis Library (companion to~\cite{giannakopoulos2014introduction}) to perform signal processing on audio clips collected by GVA. In total, we developed four classifiers targeting different aspects of the audio signal. These classifiers are described below.

\vspace{2mm}

\noindent \textbf{Noise Detection:} In their study,~\cite{malkin2019attitudes} found that 2.93\% of all audio clips were noise/gibberish, suggesting accidental triggering. Another 1.72\% of all audio clips classified as accidental recordings contained pre-recorded audio, which could also be noisy if the primary audio source was further from the recording device or located in a noisy environment. We used Signal to Noise Ratio (SNR) as the metric to detect such audio clips. For each audio clip, we first used Spectral Filtering (implemented using routines from the VOICEBOX Toolbox~\cite{brookes1997voicebox} for MATLAB) to separate the background noise from the main contents of the file and then used these two signals to compute the SNR. All audio clips that were having SNR$<$0 were classified as possibly sensitive.

\vspace{2mm}

\noindent \textbf{Speaker Diarization:} In their study,~\cite{malkin2019attitudes} found respondents to be more protective of others' privacy, feeling more uncomfortable with the smart speaker storing voices of other people, such as their children. We used speaker diarization to detect audio clips having multiple speakers and/or multiple pauses in an audio clip, where diarization refers to the process of partitioning audio files into homogeneous segments based on speaker identity~\cite{wiki:diarization}. More specifically, we used the speaker diarization function provided by the MATLAB Audio Analysis Library to carry out speaker diarization for each audio clip. All audio clips having two or more speaker segments were classified as possibly sensitive.

\vspace{2mm}

\noindent \textbf{Speaker Classification:} Speaker classification refers to any decision-making process that uses features of the speech signal to determine the characteristics of the speaker of a given utterance~\cite{atal1976speaker}. For this model, we specifically focused on sex classification, which is a sub-problem of speaker classification. By identifying the dominant speaker class in each audio clip, we could not only accurately predict the sex of each speaker, but also segregate all audio clips having an opposite-sex speaker as the dominant speaker. Such a case might correspond to the use of GVA by another person. For this task, we used the speaker classification function provided by MATLAB Audio Analysis Library and trained a model for the same using the Mozilla Common Voice Single Word Target Segment dataset~\cite{commonvoice}. We ran this model across all audio clips for a given user to identify the dominant speaker class i.e. male/female. All audio clips having a non-dominant speaker class as the dominant speaker were classified as possibly sensitive. Note that all background noise was separated using Spectral Filtering as mentioned earlier.

\vspace{2mm}

\noindent \textbf{Background Classification:} Background classification is another sub-problem of speaker classification. For this model, we aimed to classify the type of background noise in an audio clip into different categories such as wind, rain, ocean, highway, traffic, chatter. Files belonging to different background categories might correspond to accidental triggering, secondary data collection by intentional or unintentional recording of another person, and even a crude estimation of location. Once again, we used the MATLAB Audio Analysis Library and trained a model for the same using publicly available audio on streaming platforms such as YouTube. We ran this model across all audio clips for a given user to identify the dominant background class. All audio clips having a non-dominant background class as the dominant background were classified as possibly sensitive. Note that all audio contents were separated using Spectral Filtering as mentioned earlier.

\subsection{Natural Language Processing} We used different NLP models and techniques to identify possibly sensitive transcripts collected by GVA. In total, we developed eight classifiers, four for the transcripts having a companion audio clip and four for the transcripts that do not have a companion audio clip. These have been described below.

\vspace{2mm}

\noindent \textbf{Grammatical Error Detection:} Upon careful analysis of user transcripts during pilot runs, we found that some transcripts had grammatical errors such as jumbled up words or repetition of phrases and words. Such a case might correspond to accidental recording, as well as the use of GVA in a hurry, or by a child. We used the language-check library~\cite{languagecheck} in Python along with its available US English language dictionary to identify transcripts having one or more grammatical errors. Since most of these transcripts did not have proper punctuation, such as a `?' at the end of the command, we forcibly suppressed the error noticed by our language checker in such a case. All transcripts having one or more grammatical errors were classified as possibly sensitive.

\vspace{2mm}

\noindent \textbf{Non-standard Word Detection:} People often use voice assistants for knowledge-based queries containing uncommon words, which we refer to as non-standard words. However, such a case might correspond to an inquiry about a niche product, service, or place that one might not want others to know about. We used the pyenchant library~\cite{pyenchant} in Python along with its available US English spell checker to identify transcripts having one or more such words. All transcripts having one or more non-sensitive words were classified as possibly sensitive.

\vspace{2mm}

\noindent \textbf{Regret Word Detection:} People often use inappropriate language and engage in small talk with voice assistants, knowingly or unknowingly. Although these talks may seem fun and harmless at the time, but they might cause a serious breach of personal privacy in case an attacker gains access and misuses this data. Our model was heavily inspired by the work of Wang et al. and Zhou et al., who performed qualitative studies to identify regret words on Facebook and Twitter, respectively.~\cite{yang2011regret, zhou2016regret}. We adopted the category-wise comprehensive list of regret lexicons collected by~\cite{zhou2016regret} for our study. To identify transcripts having such regret words, we performed basic NLP preprocessing such as tokenization and stemming and then did a keyword match with the database of regret words. All transcripts having one or more regret words were classified
as possibly sensitive.

\vspace{2mm}

\noindent \textbf{Negative Sentiment Detection:} Just like the previous model, this model was motivated by people’s tendency to engage in small talk with voice assistants. However, sometimes these transcripts (and associated audio clips) might convey negative emotions such as sadness, hate, discrimination, etc., that one might not want others to know about. We decided to add a separate model for negative sentiments apart from the regret word detection model to capture commands such as `Stop' or `Shut up' that do not contain any regret word per se but might be sensitive to some people. We used the NLTK library~\cite{bird2009natural} in Python along with its available VADER sentiment intensity analyzer to identify transcripts having a negative sentiment. All transcripts having a negative sentiment were classified as possibly sensitive.

\section{Feature Selection}
\label{subsec:features}

\noindent We included audio-based, text-based, and user-based features to obtain the final feature vectors in our training dataset. Each feature has been briefly explained below.

\subsection{Audio-based features}

\noindent \textbf{MFCC:} The Mel-Frequency Cepstrum (MFC) represents the short-term power spectrum of a sound. Mel-Frequency Cepstral Coefficients or MFCCs are coefficients that collectively make up this representation. We use the mean and standard deviation values for the first 13 coefficients as a feature vector of length 26.

\vspace{2mm}
\noindent \textbf{Spectral contrast:} Spectral contrast extracts spectral peaks, valleys, and their differences in each sub-band, to represent the relative spectral characteristics of an audio clip. We use it as a feature vector of length 7.

\vspace{2mm}
\noindent \textbf{Tempo:} Tempo is the pace or speed at which a section of music is played. We use it as a feature vector of length 1.

\vspace{2mm}
\noindent \textbf{SoundNet:} We use sound representations from the last layer of the SoundNet network~\cite{aytar2016soundnet} as a feature vector of length 1024.

\subsection{Text-based features}

\noindent \textbf{LIWC:} We use the 64 semantic categories in the Linguistic Inquiry and Word Count~\cite{pennebaker2001linguistic} dictionary to construct a feature vector of length 64. Each element in this vector represents a particular semantic category, its value being equal to the number of words of that category present in a transcript.

\vspace{2mm}
\noindent \textbf{Sentence embedding:} We obtain 525-dimensional spaCy word embeddings corresponding to each word in a transcript and then take their average to construct a feature vector of length 525.

\vspace{2mm}
\noindent \textbf{Swear words:} We borrow a list of swear words (such as bulls***, f******, hell, n****, queer, etc.) from~\cite{maity2018understanding} with the intuition that sensitive transcripts might contain swear words. We then construct a feature vector of length 72, where the first 71 elements indicate the presence or absence (encoded as 0 or 1) of each swear word in a transcript, and the last element indicates the total number of swear words in the transcript.

\vspace{2mm}
\noindent \textbf{Regret words:} We borrow the regret lexicon compiled by~\cite{zhou2016regret} with the intuition that sensitive transcripts might contain regret words. The lexicon is comprised of eight categories- cursing, drug, health, relationship, racial and religion, sex, violence, and work. We then construct a feature vector of length 8, where each element indicates the count of regret words from one of eight regret lexicon categories in a transcript.

\vspace{2mm}
\noindent \textbf{Sentiment:} We use the MPQA~\cite{deng2015mpqa} and NRC~\cite{mohammad2013nrc} lexicons to obtain the sentiment for each word in a transcript. This information is then used to construct a feature vector of length 4, where each element indicates the count of positive/negative sentiment words present in the transcript according to the MPQA/NRC lexicon, respectively.

\vspace{2mm}
\noindent \textbf{Emotion:} We constructed an emotion feature having 8 fields representing the count of anger, anticipation, disgust, fear, joy, sadness, surprise, and trust. This feature was constructed using the NRC emotion lexicon~\cite{mohammad2013nrc}.

\vspace{2mm}
\noindent \textbf{Unigrams \& Bigrams:} We obtain the top 100 unigrams and top 100 bigrams in our dataset and construct two feature vectors of length 100 corresponding to both unigrams and bigrams. Each element in these feature vectors indicates the count of a top 100 unigram/bigram in a transcript.

\subsection{User-based features}

\noindent \textbf{Age range:} We encode participant responses to the question-\textit{What is your age?} in Survey 1 as a feature vector of length~3.

\vspace{2mm}
\noindent \textbf{Gender:} We encode participant responses to the question-\textit{With what gender do you identify?} in Survey 1 as a feature vector of length 2.

\vspace{2mm}
\noindent \textbf{Age of Google Account:} We encode participant responses to the question-\textit{You needed a Google account to log into Google services on your current Android smartphone. How long have you been using that Google account on Android smartphones?} in Survey 1 as a feature vector of length 3.

\vspace{2mm}
\noindent \textbf{Frequency of GVA usage:} We encode participant responses to the question-\textit{How often do you use Google Assistant on your Android smartphone?} in Survey 1 as a feature vector of length 3.

\vspace{2mm}
\noindent \textbf{Span of GVA usage:} We encode participant responses to the question-\textit{Since how long have you been using Google Assistant on your Android smartphone?} in Survey 1 as a feature vector of length 3.

\vspace{2mm}
\noindent \textbf{Association with CS or a related field:} We encode participant responses to the question-\textit{Are you majoring in or do you have a degree or job in computer science, computer engineering, information technology, or a related field?} in Survey 1 as a feature vector of length 2.

\end{document}